\documentclass{elsarticle}
\pdfoutput=1
\usepackage{amssymb}
\usepackage{amsmath}
\usepackage{amsthm}
\usepackage{subcaption}
\usepackage{graphicx}
\usepackage{hyperref}
\usepackage{soul}
\usepackage{float}
\usepackage[cal=boondox]{mathalpha}
\pdfmapfile{=boondox.map}
\usepackage[margin=2.0cm]{geometry}
\usepackage{array}
\usepackage{booktabs}
\usepackage{color}
\usepackage{pdflscape}
\usepackage{xargs}
\usepackage{bm}
\usepackage{url}
\usepackage{lineno}
\usepackage[colorinlistoftodos]{todonotes}

\newcommandx{\unsure}[2][1=]{\todo[linecolor=red,backgroundcolor=red!25, bordercolor=red,#1]{\tiny{#2}}}
\newcommand{\figref}[1]{\hyperref[#1]{Fig. \ref*{#1}}}
\newcommand{\tabref}[1]{\hyperref[#1]{Table \ref*{#1}}}
\newcommand{\secref}[1]{\hyperref[#1]{Section \ref*{#1}}}
\newfloat{gallery}{thp}{lob}
\newcolumntype{M}[1]{>{\centering\arraybackslash}m{#1}}
\newtheorem*{claim}{Claim}

\title{Sentinel-1 Additive Noise Removal from Cross-Polarization Extra-Wide TOPSAR
  with Dynamic Least-Squares}

\begin{document}
\begin{frontmatter}
  \author[syde]{Peter Q. Lee}
  \author[syde]{Linlin Xu}
  \author[syde]{David A. Clausi}
  
  \address[syde]{Department of Systems Design Engineering, Faculty of
    Engineering, University of Waterloo, 200 University Avenue West
     Ontario, Canada, N2L 3G1}

\pagenumbering{arabic}

\begin{abstract}
Sentinel-1 is a synthetic aperture radar (SAR) platform with an operational mode called extra wide (EW) that offers large regions of ocean
areas to be observed. A major issue with EW images is that the
cross-polarized HV and VH channels have prominent additive noise
patterns relative to low backscatter intensity, which disrupts tasks that require manual or automated interpretation. The European Space Agency (ESA) provides a
  method for removing the additive noise pattern by means of lookup
  tables, but applying them directly produces unsatisfactory
results because characteristics of the noise still remain. 
Furthermore, evidence suggests that the magnitude of the additive noise dynamically depends on factors that are not considered by the ESA estimated noise field.

To address these issues we propose a quadratic objective function
to model the mis-scale of the provided noise field on an image. We
  consider a linear denoising model that re-scales the noise field for
  each subswath, whose parameters are found from a least-squares
  solution over the objective function. This method greatly
reduces the presence of additive noise while not requiring a set of
training images, is robust to
heterogeneity in images, dynamically estimates parameters for each image, and finds parameters using a closed-form solution.

Two experiments were performed to validate the proposed method.
The first experiment simulated noise removal on a set of RADARSAT-2
images with noise fields artificially imposed on them. The second
experiment conducted noise removal on a set of Sentinel-1 images taken
over the five oceans. Afterwards, quality of the noise removal was
  evaluated based on the appearance of open-water. The two
experiments indicate that the proposed method marks an improvement
  both visually and through numerical measures. \footnote{This manuscript has been accepted in Remote Sensing of Environment, whose version of record can be found at \url{https://doi.org/10.1016/j.rse.2020.111982}. \textcopyright 2020. This manuscript version is made available under the CC-BY-NC-ND 4.0 license \url{http://creativecommons.org/licenses/by-nc-nd/4.0/}}

\end{abstract}

\begin{keyword}
  Sentinel-1, SAR, Noise removal, Denoising, Additive noise, Thermal noise, Banding, Scalloping
\end{keyword}

\end{frontmatter}
\section{Introduction}

Synthetic aperture radar (SAR) is method of remote sensing that is
useful for monitoring the surface of the planet while being insensitive to atmospheric conditions. Applications include topographical mapping through SAR interferometry \citep{land,pepe2017review}, sea ice mapping/classification
\citep{fcn_scott, tanice, sent1autoseaice, linlin_seaice, irgs_paper},
oil spill detection \citep{oilspill, topouzelis2008oil}, ship detection
\citep{ship, ship2}, and others \citep{sar_veg_applications, general_sar_applications}. Consequently, SAR imagery is an essential tool for
global monitoring. The Sentinel-1 program, which was created and is administered by
the European Space Agency (ESA), operates two SAR satellites,
Sentinel-1A (launched 2014) and Sentinel-1B (launched 2016), and acquires and distributes SAR images
taken over Earth. The
diverse set of applications for SAR and the fact that the Sentinel program provides open-data solidifies the potential for research and commercial
products.

Sentinel-1 operates in a number of acquisition modes. This
paper focuses on the Sentinel-1 extra-wide (EW) mode, which is mainly
used over ocean regions, with the images acquired using a method called TOPSAR \citep{topsar}. Sentinel-1 satellites emit bursts of horizontally or vertically polarized radiation and receives polarized backscatter to form co-polarized
images, where the polarization of the emission is the same as the
polarization of the received signal, or cross-polarized images, where the
polarization of the emission is orthogonal to the polarization of the
received signal. A caveat of EW mode for Sentinel-1 is that there are
prominent additive noise patterns present in the cross polarization
channels with respect to lower backscatter intensity \citep{esathermal, thermal, tanice}. These are sometimes called banding or scalloping effects in the
literature \citep{fcn_scott, bandinscansar}. These noise patterns can cause major issues for
interpretation-based
tasks because the added intensity corrupts the true backscattered
signal of the target \citep{esathermal}. ESA currently
provides lookup tables to estimate the additive noise in terms of azimuth and range components
as detailed in their Instrument Processing Facility (IPF) product \citep{esathermal}, and
disseminated through noise calibration \textit{eXtensible Markup
  Language} (XML) files that are
distributed with each Sentinel-1 scene.  By interpolating the lookup
tables, an estimated additive noise field can be derived for the
original image. This paper focuses
  specifically on data using IPF version 2.91, however as of writing
  IPF version 3.20 is the latest version that presents improvements in
computing the additive noise field. Working with data from IPF
versions less than or equal to 2.91 is problematic, however, as
subtracting the noise field from the image directly is insufficient as shown by others \citep{thermal, tanice, baltic} and ourselves in
\secref{sec:method}. The problem is severe enough that some practitioners prefer to discard the first subswath to make modeling easier \citep{tanice, sent1autoseaice}.

While there is a significant body of work aiming to reduce speckle noise \citep{gagnon1997speckle,
  parrilli2011nonlocal, qiu2004speckle, 5958515,
  santoso2015performance, 8460380, 7053819}, there are considerably fewer published methods on reducing the additive noise in SAR images.

 \cite{kalman_scansar} modeled the additive noise field of ScanSAR images with gain and offset parameters and implemented a Kalman filter to estimate the parameters for the azimuth and range in a decoupled manner. However, it is not immediately clear how this could be extended to Sentinel-1 EW where the magnitude of the additive noise varies greatly between adjacent subswaths. Specific to Sentinel-1 EW, \cite{baltic} used an approach with a
finite impulse response filter to reduce the scalloping in the azimuth
direction selectively over ocean regions. However, this has the computationally expensive requirement of computing texture
features, with the possibility of incorrectly modifying the intensity of the intended target within
the image. \cite{thermal} created an approach for Sentinel-1 EW that modifies 
the estimated noise field with linear scaling and intercept parameters in an attempt to
eliminate the additive noise in the image. The values of the
parameters were estimated with an iterative grid search regression approach using a
large sample of Sentinel-1 ocean scenes, with the mean-estimates of
the parameters used for evaluation. 
A few drawbacks of this approach include the requirement of selecting samples of
images across uniform ocean areas, a priori. The method also makes the assumption that the
ideal scaling parameters are the same for every scene, an assumption
that we found to not always be valid, with an example of this shown in \figref{fig:contradiction}. Finally, if one wanted to adapt the method
to dynamically estimate the parameters for individual scenes, it would be limited because of the
aforementioned uniform ocean requirement.

The method proposed in this paper builds upon the approach of scaling the noise field for each
subswath as proposed by \cite{thermal}. Our contribution is based both on
creating an objective function that incorporates the
characteristics of the additive noise and simultaneously choosing
scaling parameters for each subswath to minimize the objective
function. Our method leads to four main aspects of improvement:
\begin{enumerate}
  \item Parameters may be estimated on images with heterogeneous
    features (e.g. containing open-water and sea-ice, cyclones, etc.).
  \item No training set of images is needed; the method can be applied for each
    individual image without prior knowledge of the scenes.
  \item Parameters are dynamically estimated for each image. This
    is significant because the ideal scaling is different for each
    image, which makes using static parameter estimates suboptimal.
  \item The method has a closed-form solution that can be solved
    exactly without iterative methods.
  \end{enumerate}
  Ultimately our approach reduces to estimating the parameters of a
  linear function based on a quadratic objective function using a
  least-squares solution.

Two experiments were performed to evaluate the effectiveness of the
proposed denoising method. The first experiment simulated denoising by
applying randomly scaled noise fields to non-SAR Sentinel-2 images and
then denoising them with the proposed method. The second experiment
used a sample of HV and VH polarized Sentinel-1 EW images from all
five oceans and evaluated the proposed method in terms of the flatness
of the result over selected subregions that were expected to be
flat. These experiments ultimately indicated that the proposed method
was a significant improvement over the baseline method provided by ESA
and also that estimating parameters for each image independently is
beneficial compared to a static estimation approach. We also
  perform a small experiment on data processed with an IPF above 3 that
  resulted in some modest improvement.
Therefore, this work is useful 
for any application that uses EW cross-polarized Sentinel-1,
such as ice analysis, as the method provides significantly better
denoising and does not require expensive computation or large amounts
of prior information or training sets.

\section{SAR noise background}

Before details on the mechanics of the noise are provided, a brief
background of TOPSAR will first be given. TOPSAR is a type of SAR
acquisition that captures a wide swath area 
without the typical loss of resolution or scalloping \citep{topsar}. The method
works by simultaneously acquiring subswaths that cover different
elevation angles (the angle between the ground target and the satellite) by steering the antenna aperture to focus at
different elevation angles while continuously rotating along the azimuth angle. Sentinel-1 incorporates TOPSAR in two modes,
interferometric wide (IW) swath mode, and extra wide (EW) swath mode. EW mode has subswaths enumerated as EW1,
EW2, EW3, EW4, and EW5 that are ordered by increasing elevation angle along the range. During the observation period, the satellite sweeps a
number of radar bursts along the azimuth and observes the backscattered radiation for
each of the five subswaths. The received signals
from each of the bursts are stitched along the azimuth direction (the
direction of the satellite's orbital velocity) to form the rows of the
image, while the signal corresponding to the five subswaths along the
range (the direction perpendicular to the satellite's orbital velocity and
tangent to the Earth's surface) are stitched together to form the
columns of the image. In the supplementary materials we provide an animation that approximates the TOPSAR process.

ESA provides a forward model for estimating the additive noise induced by
TOPSAR as a function of a number of factors 
\begin{equation}
  y_a(i,j) = \frac{D(i) P(a,i)}{E(a,i,j) R(j)} f,
\end{equation}
 for azimuth index $i$, range index $j$ that are within a subswath $a$ \citep{algorithm}. $P(a,i)$ is the power gain term that is used for drift correction while orbiting. $R(j)$ is the range spreading loss term used to
 correct errors from the range compression algorithm. Scalar $f$ is calibration parameter determined by the noise and the processor. The
 remaining two terms are based on the radiation
 pattern of the antennas, with
 $D(i)$, the descalloping gain, being inversely proportional to the antenna array pattern with
 respect to the azimuth, and $E(a,i,j)$, the elevation antenna pattern gain, being proportional to the
 antenna pattern with respect to the range. The reader is referred to
 \cite{algorithm} for additional details. In antenna theory, the power of the radiation pattern for an antenna is directly related to the strength of the received signal at the corresponding angle from the receiving antenna \citep{antenna}.
Consequently, the radiation pattern is a fundamental
variable for the shape of the additive noise.

\begin{figure}[H]
\begin{subfigure}[t]{0.48\linewidth}
  \includegraphics[width=\linewidth]{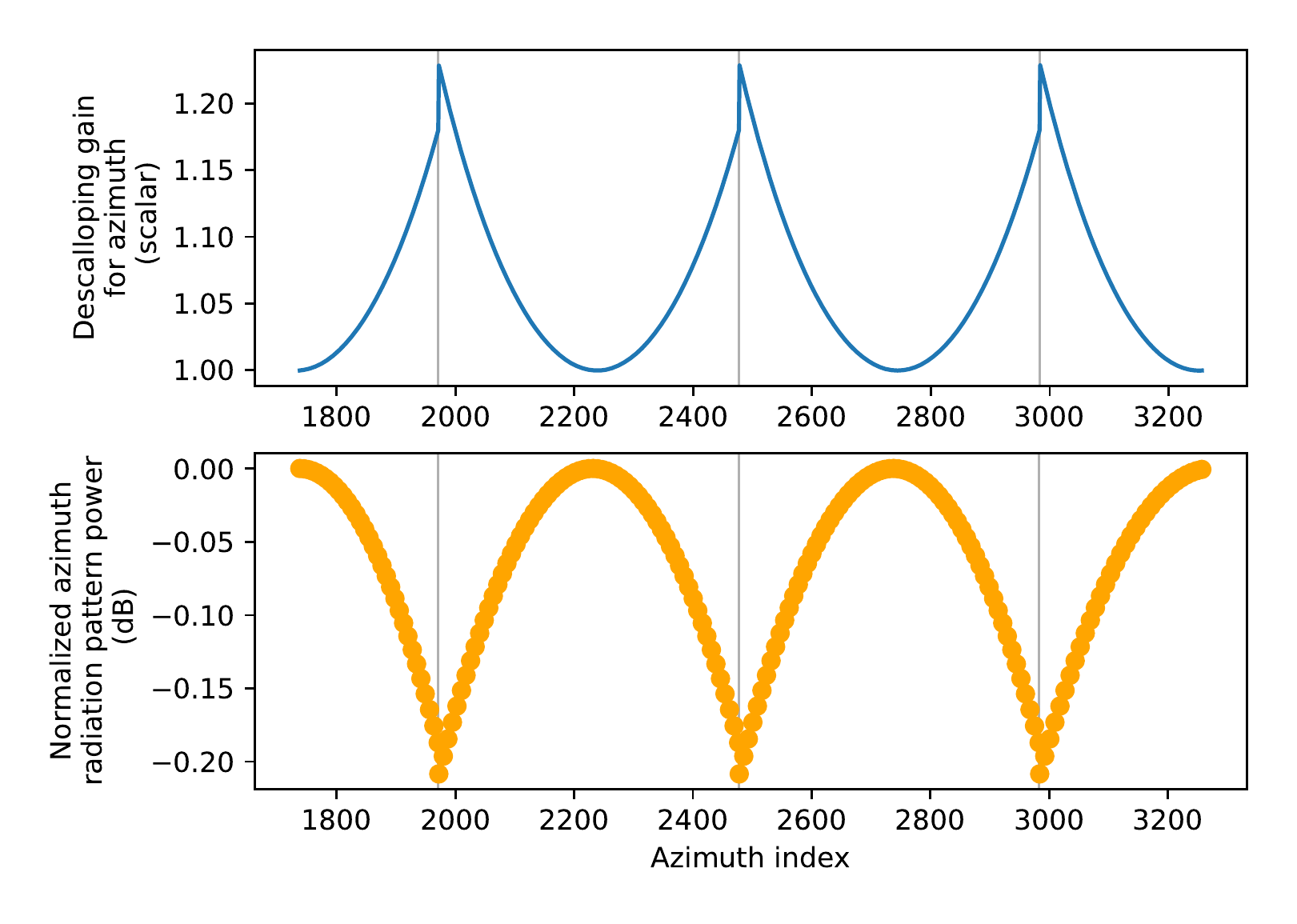}
  \caption{Top: The estimated noise contribution in the azimuth
    direction (descalloping gain).\\
    Bottom: Average radiation pattern power gain with respect to azimuth.\\
    The vertical demarcations indicate the beginning and end of
    observations obtained from a burst. Only a portion of the azimuth is shown for brevity.}
  \label{fig:az_radiation_vs_gain}
\end{subfigure}\hfill
\begin{subfigure}[t]{0.48\linewidth}
  \includegraphics[width=\linewidth]{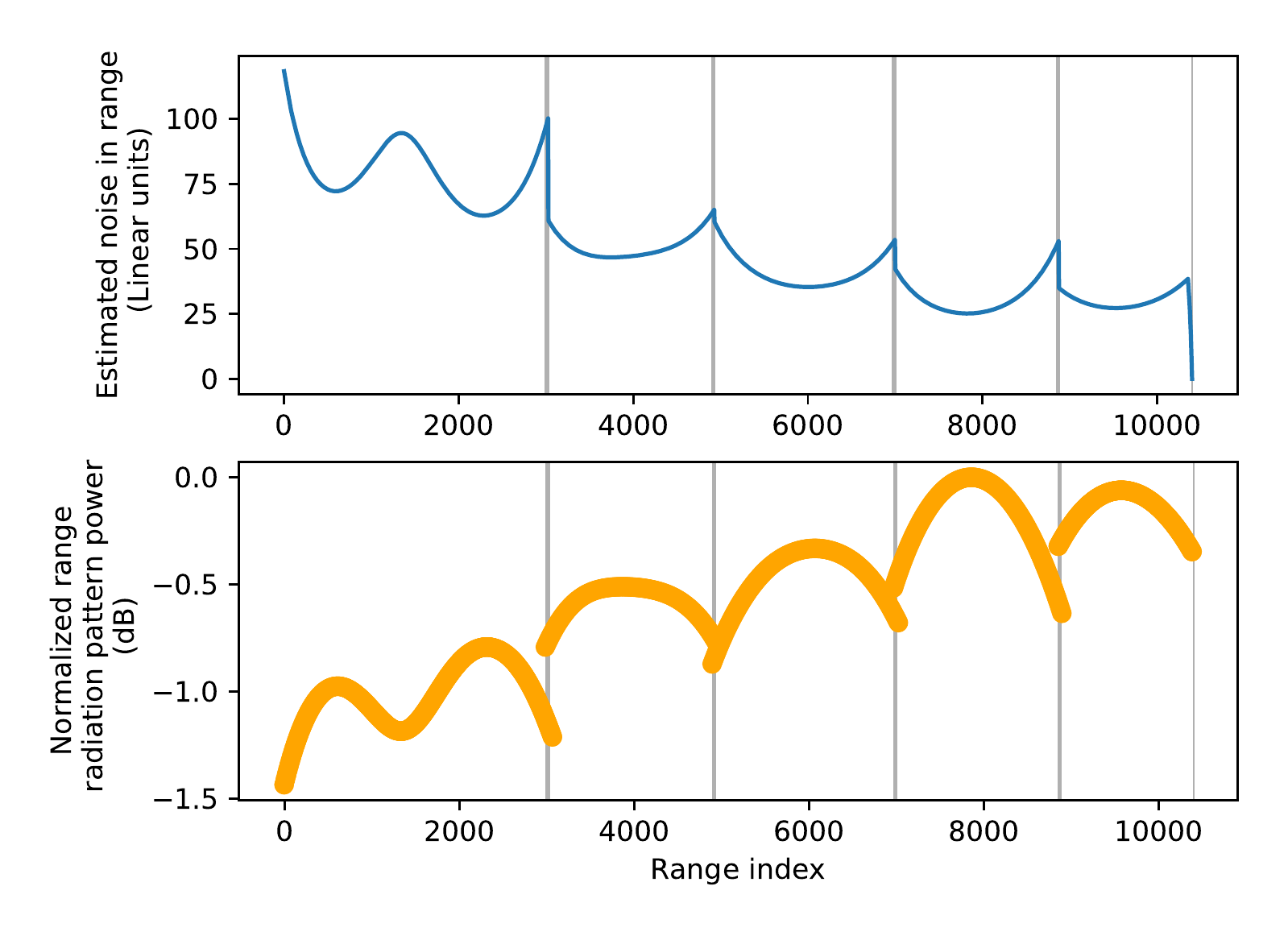}
  \caption{Top: The estimated noise contribution in the range
    direction.\\
    Bottom: Radiation pattern power gain with respect to range.\\
    The vertical demarcations indicate the beginning and end of subswaths.
  }
  \label{fig:radiation_vs_gain}
\end{subfigure}
\caption{Comparison between the estimated noise contributed by the
  azimuth and range with respect to radiation pattern for EW TOPSAR.
  The measurements for the antenna
  pattern were initially given
in angles and were converted to the corresponding range and azimuth
indices in order to show the negative relationship between noise and
radiation pattern. The measurements correspond to the
noise field as shown in \figref{fig:noise_construction}.}
\end{figure}

As Sentinel-1 uses an array antenna, the observed signal in the azimuth direction will naturally be an aggregate function of the radiation patterns for each element in the array as they are
rotated along the azimuth
\citep{algorithm, antenna}. \figref{fig:az_radiation_vs_gain} shows an example of the relationship between the 
average antenna element pattern for the burst along the azimuth, derived from the Sentinel-1
calibration XML file, and azimuth component of the estimated noise.
As the bursts are stitched along the azimuth \citep{topsar}, the aggregate pattern appears as U-shaped and symmetric. The corresponding angles closer to
the burst centre have higher gain, which reduces the magnitude of
additive noise compensation needed. Angles further from the
centre have less antenna-gain
and require more noise compensation.

The negative relationship between the noise and radiation
pattern is apparent in the range direction as
well. \figref{fig:radiation_vs_gain} shows a comparison between the range noise and
the radiation pattern for each subswath. When the antenna is steered to
acquire a different elevation angle, and hence different subswath, the subsequent radiation pattern changes. This is most notable in the
first subswath, where the radiation pattern has two main lobes (local
maxima of the gain). The remaining subswaths only have one main lobe,
but have different relative magnitudes of radiation power.

Within the noise calibration XML
files in every Sentinel-1 EW product, as of IPF 2.9, the information to
compute the noise field $y_a(i,j)$
is provided within two different lookup tables labelled as the
\textit{noiseRangeVector},
which models $y_a(i,j)/D(i)$, and the
\textit{noiseAzimuthVector}, which models
$D(i)$. By performing linear interpolation between entries in the lookup
tables and multiplying the two results together, the estimated noise
field $y_a(i,j)$ can be constructed. Given $x$ as the square of the
digital pixel values in the original image, ESA recommends
denoising the measurements by subtracting the noise field from the image
\begin{equation}
  \phi_a(i,j) = x_a(i,j) - y_a(i,j),
  \label{eq:esa_model}
\end{equation}
where the square root of $\phi$ returns the result to the original linear units / digital values \citep{esathermal}. As this step is performed prior to any other calibration procedures, which differ based on application, we will continue to operate in terms of linear units throughout this paper. Further, for a given subswath $a \in \mathcal{A}=\{\text{EW1}, \text{EW2}, \text{EW3}, \text{EW4}, \text{EW5}\}$, let the
azimuth row and range column be $(i,j) \in a$. Then \ref{eq:esa_model} is represented more succinctly as
\begin{equation}
  \phi_a = x_a - y_a.
  \label{eq:esa_condensed}
\end{equation}

\begin{figure}[H]
  \includegraphics[width=\columnwidth]{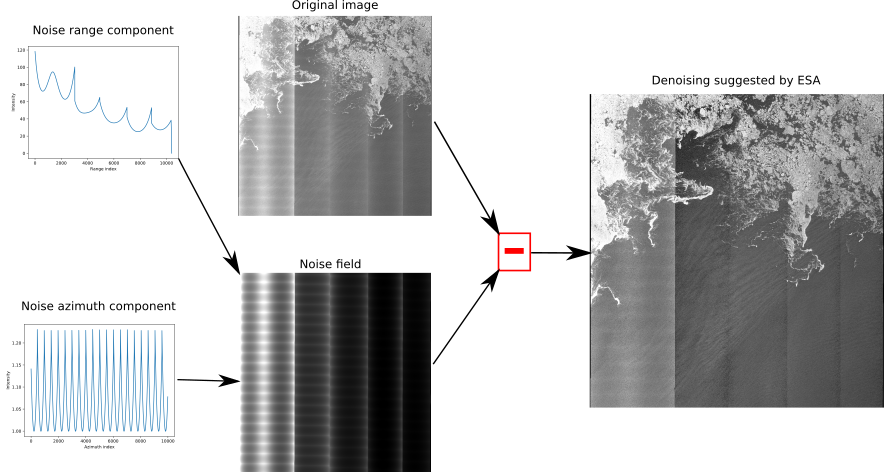}
    \caption{Visualization of the construction and application of the
      noise field using the noise vectors. Overall, the final
      denoising is unsatisfactory, most prominently in EW1 where
      patterns of the noise field are blatant in the result image.}\label{fig:noise_construction}
\end{figure}

\begin{figure}[H]
  \begin{subfigure}[t]{0.48\linewidth}
    \includegraphics[width=\linewidth]{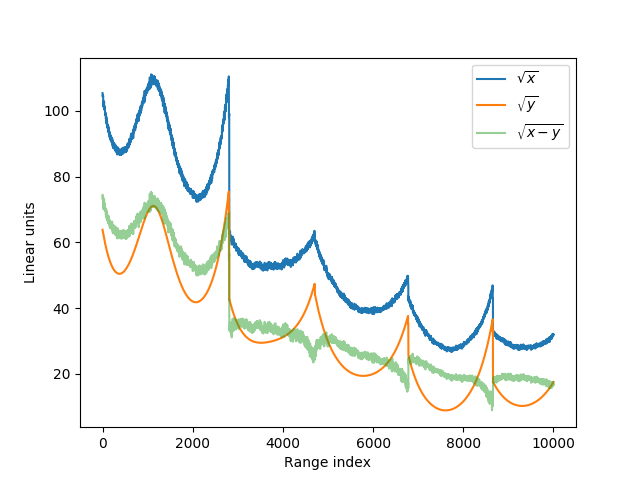}
    \caption{Mean intensity of measurement, noise, and their difference
      over ocean area with respect to \textbf{Range}.}
  \end{subfigure}\hfill
  \begin{subfigure}[t]{0.48\linewidth}
    \includegraphics[width=\linewidth]{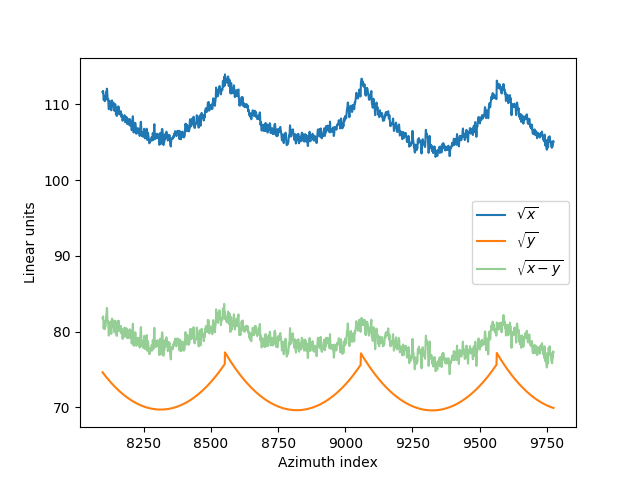}
    \caption{Mean intensity of measurement, noise, and their difference
      over ocean area with respect to \textbf{Azimuth} for EW1.}
  \end{subfigure}
  \caption{Plots of mean intensities of measurements in sections of ocean with respect
    to each direction. Ideally the subtraction would
    produce a flat profile, but aspects of the noise are still present
    in both directions.}\label{fig:line_unscaled}
\end{figure}

However, applying (\ref{eq:esa_condensed}) directly is 
insufficient. \figref{fig:noise_construction} provides a
visualization of the process of constructing the noise field and the
result of subtracting it from the original image.
The subtraction results in subswaths being under or over
compensated. For example, \figref{fig:line_unscaled} shows the values
along the azimuth and range directions of an open-water area, where the result has noise
patterns still present. This implies that the strength of the
noise field needs to be scaled differently for each subswath. \cite{thermal}
proposed a method that chooses vectors for scaling coefficients $\mathbf{\bar{k}} =
[\bar{k}_{\text{EW1}}, \bar{k}_{\text{EW2}}, \bar{k}_{\text{EW3}}, \bar{k}_{\text{EW4}}, \bar{k}_{\text{EW5}}]$ and intercepts $\mathbf{\bar{o}} = [\bar{o}_{\text{EW1}}, \bar{o}_{\text{EW2}},
\bar{o}_{\text{EW3}}, \bar{o}_{\text{EW4}}, \bar{o}_{\text{EW5}}]$ such that 
\begin{equation}
  \bar{\phi}_{a} = x_{a} - \bar{k}_{a}y_{a} + \bar{o}_a
  \label{eq:2018_model}
\end{equation}
where $\mathbf{\bar{k}}$ and $\mathbf{\bar{o}}$ were statically
estimated using a training set of several hundred images.

\begin{figure}[H]
  \begin{subfigure}{0.5\linewidth}
    \includegraphics[width=\linewidth]{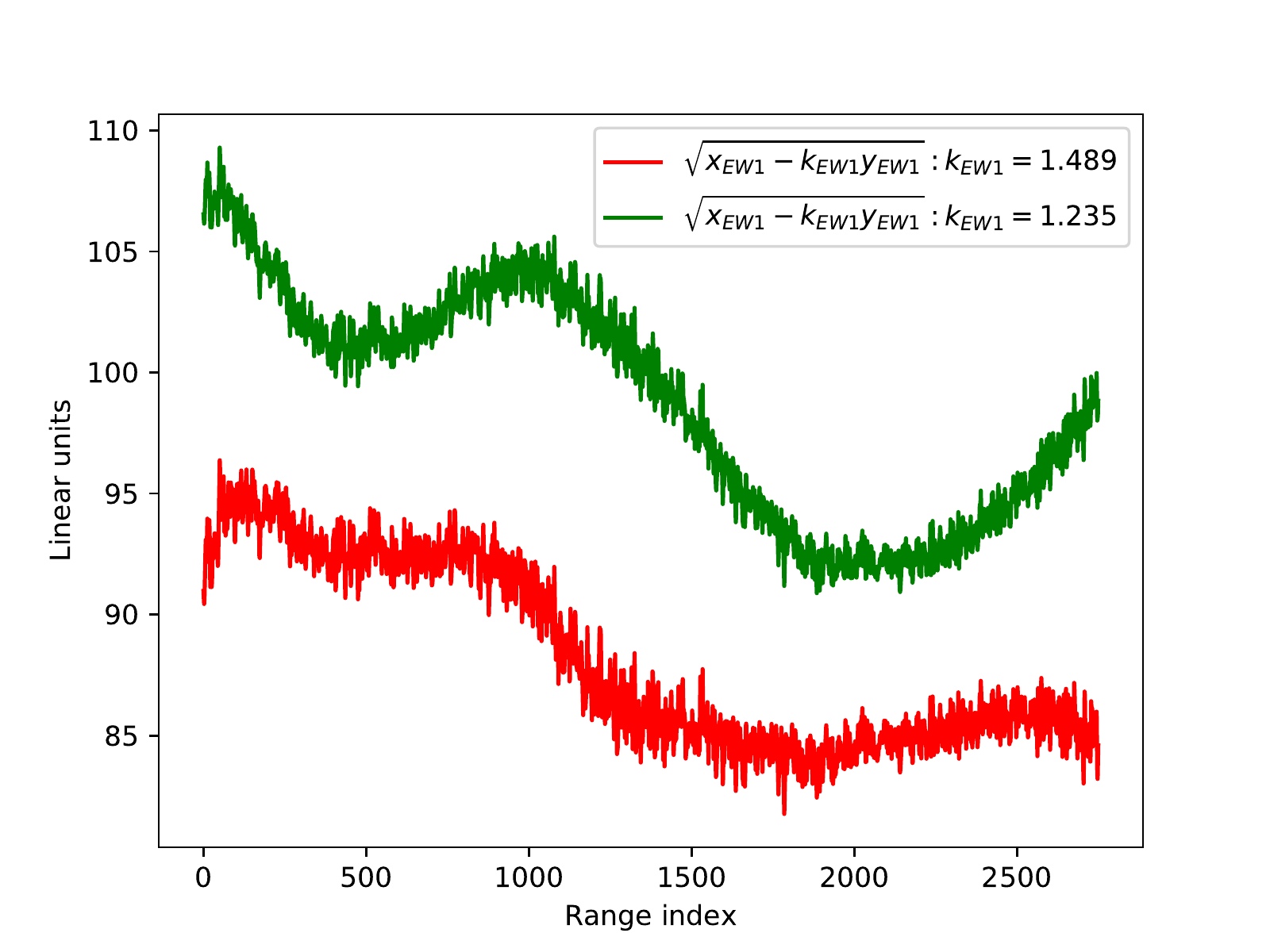}
    \caption{Scene with a higher magnitude of additive noise.}
    \label{subfig:ocean5}
  \end{subfigure}
  \begin{subfigure}{0.5\linewidth}
    \includegraphics[width=\linewidth]{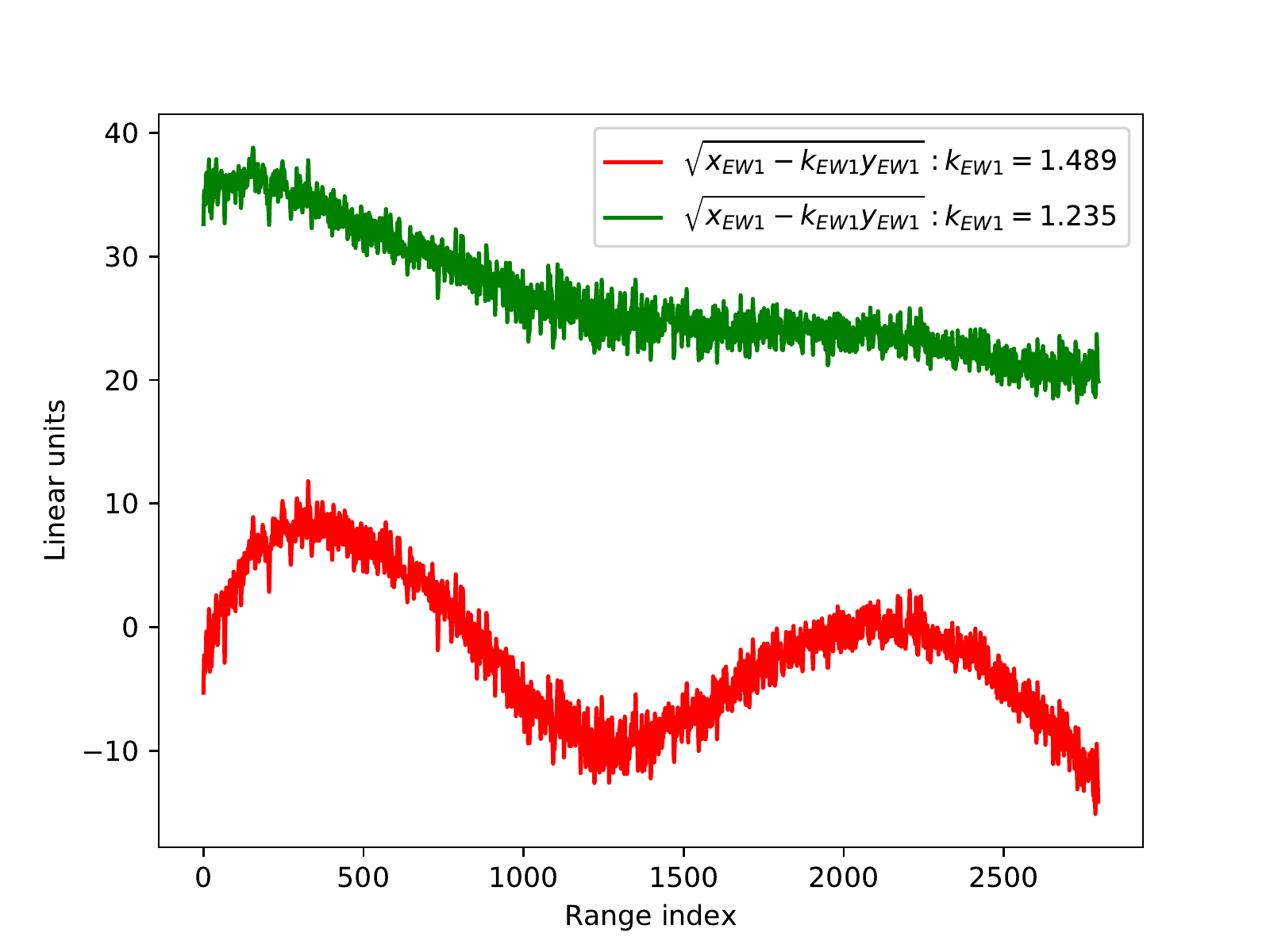}
    \caption{Scene with a lower magnitude of additive noise.}
    \label{subfig:ocean28}
  \end{subfigure}
  \caption{A comparison of the mean measurement of EW1 over ocean
    regions of two different scenes after noise removal with two
    different scaling factors ($k$). At $k=1.2350$ the left scene has
    noise patterns present while the right scene has a flat
    profile. However, at $k=1.4886$ the left scene as the noise
    patterns better compensated, but results in the right scene being overcompensated. This indicates that no single scaling factor will fit every scene.}
  \label{fig:contradiction}
\end{figure}

Another element to consider is whether the ideal scaling parameters are dependent on factors other than the sensor.
\figref{fig:contradiction} shows
that single values of $k$ do not universally fit between different images. Therefore, it would be ideal for scaling factors to be
estimated dynamically for each image.

\section{Methods}
\label{sec:method}
The goal of this work is to estimate
appropriate scaling parameters for each image individually. We make a
similar assumption to \cite{thermal} that the estimated noise in each
subswath needs to be linearly re-scaled. Our dynamic model uses a set of dynamically estimated scaling
parameters $\mathbf{\hat{k}} = [\hat{k}_{\text{EW1}}, \hat{k}_{\text{EW2}}, \hat{k}_{\text{EW3}},
\hat{k}_{\text{EW4}}, \hat{k}_{\text{EW5}}]$ such that 
\begin{equation}
  \hat{\phi}_{a} = x_{a} - \hat{k}_{a}y_{a}
  \label{eq:cur_model}
\end{equation}
Our model is distinct from (\ref{eq:2018_model}) because the scaling
parameters, $\mathbf{\hat{k}}$, are estimated for each image
independently. We also considered using a set of intercepts
$\mathbf{\hat{o}}$ but found that including these did not make a
significant impact on the end results. To estimate parameters for each image independently, an objective function, $L$, is defined based on the
characteristics of the estimated additive noise present within each image. The
values of $\mathbf{\hat{k}}$ are thereby chosen to minimize $L$
such that $\mathbf{\hat{k}} = \underset{\mathbf{\hat{k}}}
{\mathrm{argmin}}\ L$. 

\begin{figure}[H]
  \includegraphics[width=0.95\linewidth]{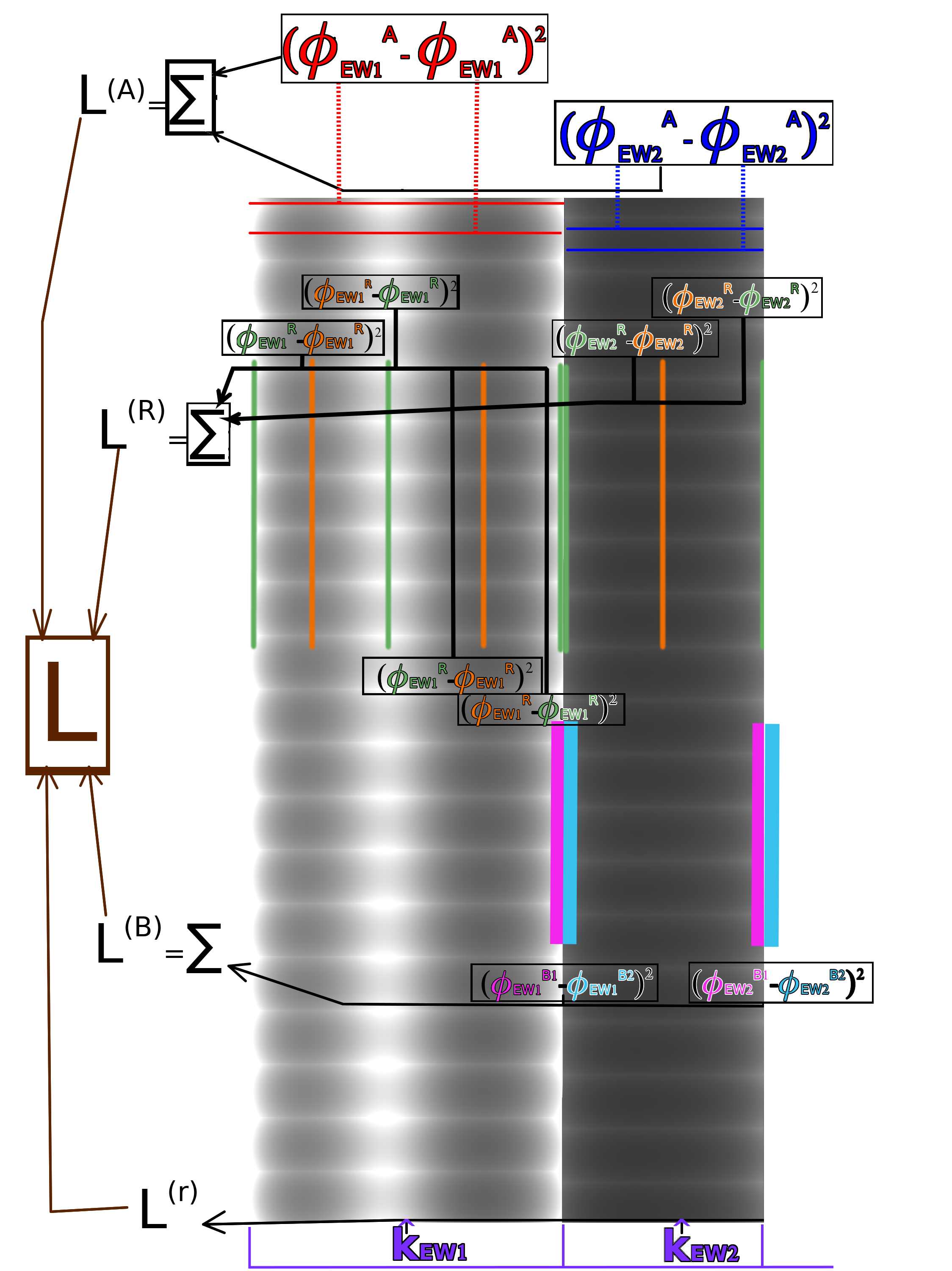}
  \caption{High level overview of the objective function for the first two subswaths, for simplicity. The noise
    field is overlaid with abridged annotations relating to the azimuth
    noise loss ($L^{A}$), range noise loss ($L^{R}$ and
    $L^{B}$), and regularization loss ($L^{r}$). While the annotations are overlaid on
    the noise field to convey the indices of the points used in the terms, note that the loss is
    a function of the measured SAR image, the estimated noise field,
    and the scaling parameters. Best viewed in colour.}
\label{fig:executive}
\end{figure}

Throughout the remainder of this section we describe the main components of the objective function (L) based on the noise characteristics in both the azimuth and range directions. Thus, $L$ is
defined as the sum of terms based on the noise-characteristics in the
azimuth direction ($L^{A}$), the range direction ($L^{R}$ and
$L^{B}$ that correspond to effects within and between subswaths respectively), and a term for
regularization ($L^{r}$). Each of these terms involve summing the
square difference between different pairs of samples within the
images. These terms are defined with more detail in the remainder of
this section, but \figref{fig:executive} provides a high-level visual overview of how $L$ is computed.

\subsection{Objective: Azimuth} \hypertarget{tg:obj_az}{}
\label{sec:obj_az}
A prominent attribute of the azimuth component of the noise is its
periodic pattern. As explained previously, the pattern is caused by
the U-shaped antenna pattern gain and the concatenation of bursts
along the azimuth. Thus, the period of the azimuth noise is
equivalent to the number of azimuth lines between the bursts, such that
the troughs (local minima) of the azimuth noise correspond to centre
of bursts (azimuth angle is 0) and are each one period apart. Also, since the gain
is U-shaped, each peak (local maxima) of the azimuth noise is located
half a period away from a trough.  Therefore, we assert that if the
amplitude of the noise is high, then the difference between the
adjacent peaks and troughs will be high.

\begin{figure}
  \begin{subfigure}[t]{0.48\linewidth}
    \includegraphics[width=\linewidth]{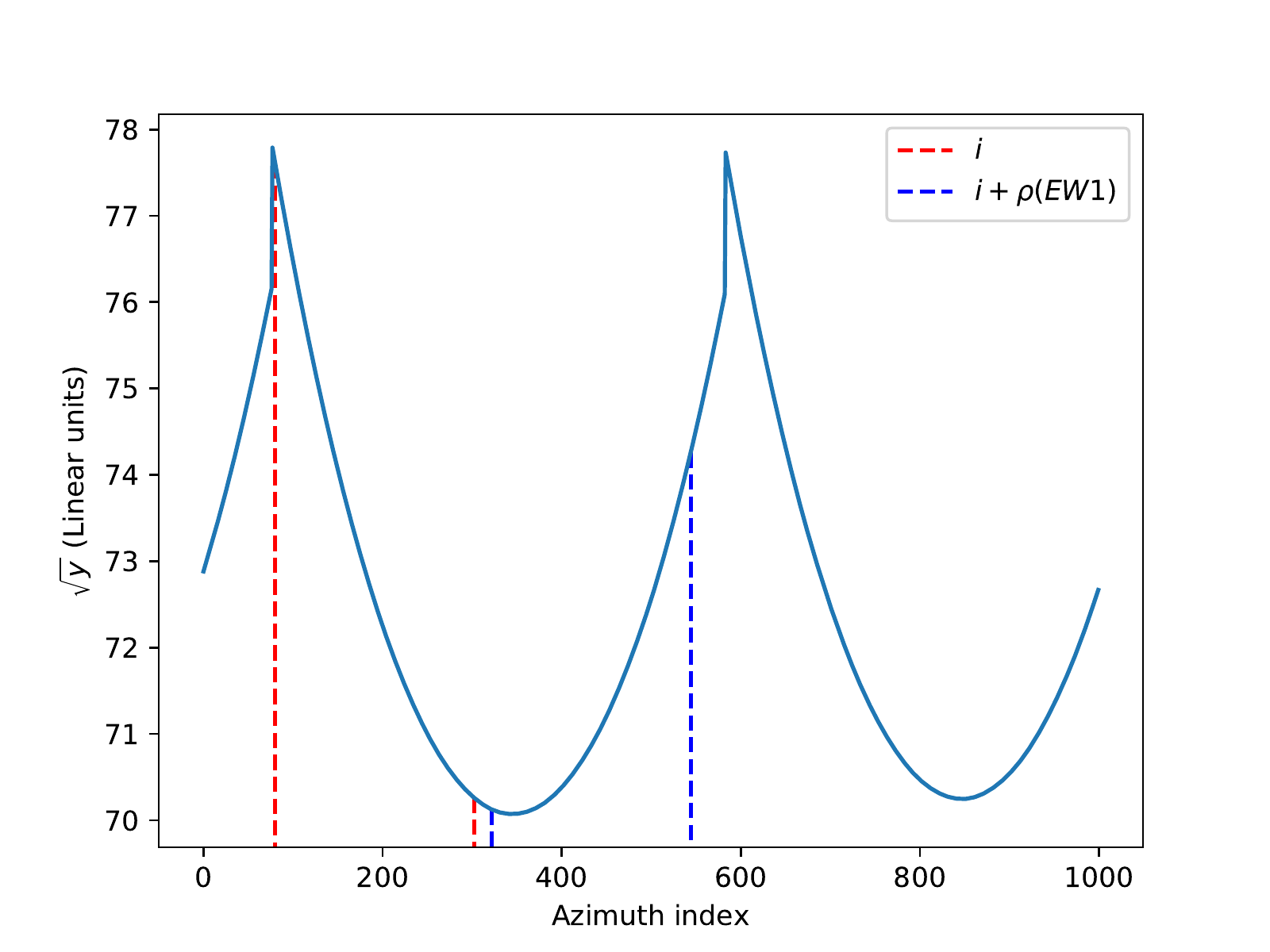}
    \caption{Noise with respect to azimuth in EW1. The dashed lines show pairs of points that are offset by half a
      burst period that are used by the objective function.}
      \label{fig:periodshift}
    \end{subfigure}\hfill
    \begin{subfigure}[t]{0.48\linewidth}
      \includegraphics[width=\linewidth]{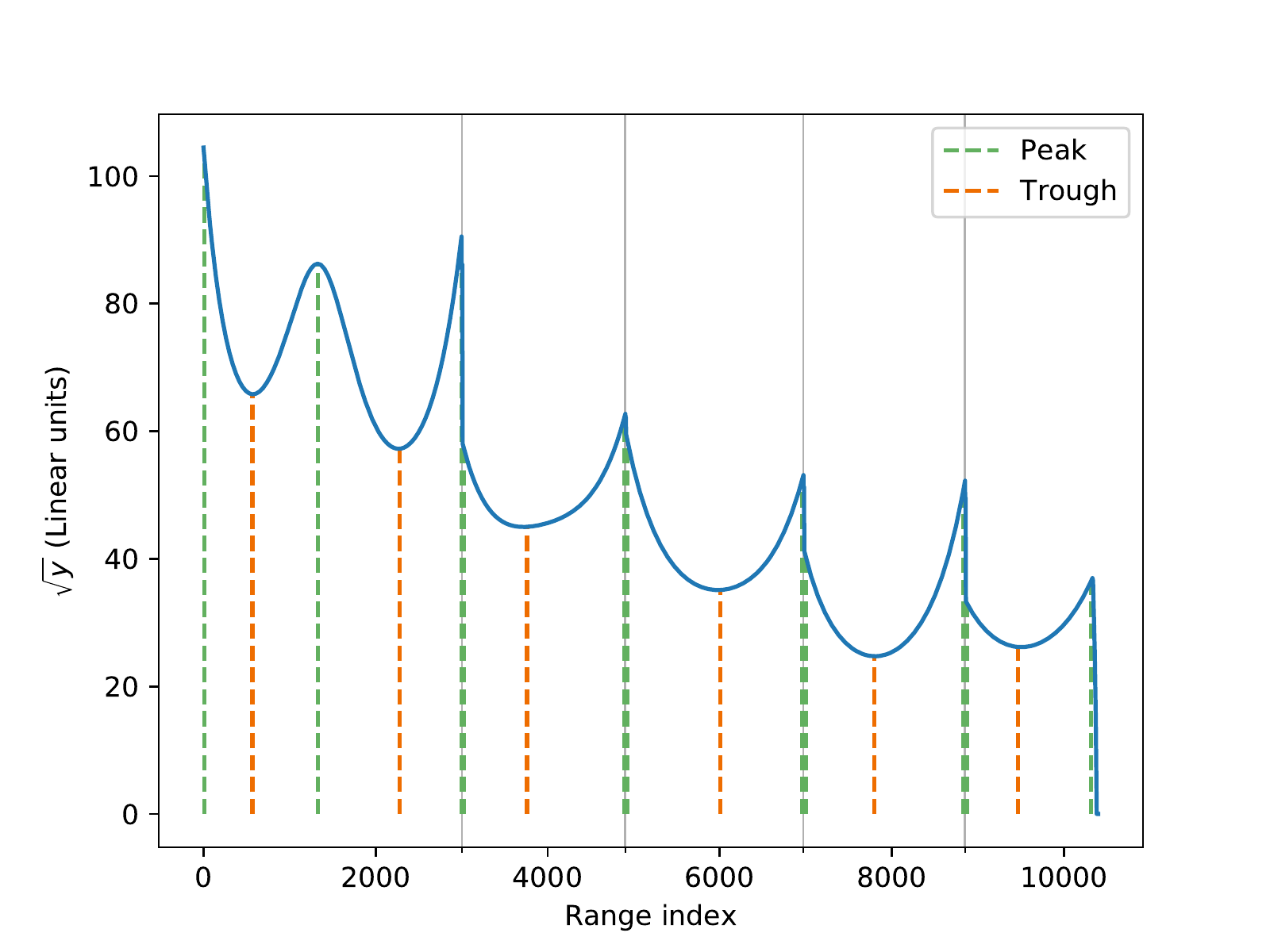}
      \caption{Range noise with peaks and troughs over all five
        subswaths. The solid lines divide the subswaths, with the
        troughs and peaks marked by the dashed lines. 
      }
  \label{fig:rg_noise}
\end{subfigure}
\caption{
  Peaks and troughs of the estimated noise with respect to azimuth and range.
}
\end{figure}

More generally, since the estimated noise within each burst is
  U-shaped and symmetric, we assume that the squared difference in the image between
  any pair of points along the azimuth will be proportional to the amplitude of the
  noise. Therefore, the pairs are selected as all possible points along the azimuth within each subswath paired with the points offset by half a period. \figref{fig:periodshift}
graphically shows this layout with respect to the azimuth.

Given the denoising model for subswath $a$ as $\hat{\phi}_a$, let
$\hat{\phi}^A_a(i)$ be the average value of all pixels for azimuth
line $i$ in $\hat{\phi}_a$. Then the azimuth component of the loss function is composed as
 \begin{equation} 
    L^{A} = \sum_{a \in \mathcal{A}}\sum_i^{N_{az}(a)}[w^{A}_a(i)[\hat{\phi}^A_a(i) - \hat{\phi}^A_a(i+\rho(a))]]^2,
    \label{eq:err_scallop}
  \end{equation}
  where $\rho(a)$ is the number of azimuth lines per
  half a burst period, computed as
  \begin{equation}
    \rho(a) = \frac{N_{burst}(a)}{2N_{az}(a)}\\
  \end{equation}
  and $N_{burst}(a)$ is the number of bursts used to construct
  subswath $a$ and $N_{az}(a)$ is the total number of azimuth lines in
  the subswath.
  Unfortunately, information containing $N_{burst}(a)$ is not
  explicitly included in the current IPF version (2.9). However, it can be
  derived as $N_{burst}(a) = N_{ap}(a) + 1$, where $N_{ap}$ is the number of \textit{antennaPattern} items for subswath $a$ within the annotation XML file. The term $w^{A}_a(i)$ is a
  weighting term introduced for subswath $a$ and azimuth line $i$
  based on the realization that the change in intensity between pairs
  of lines separated by half a burst is influenced by the backscatter of the ground targets. If the two targets have fundamentally different
  backscattering properties, for example where one line is dominated
  by ice and the other is dominated by water, then their squared
  difference will not be representative of the scale of the noise
  field in the measurement. Given $x^{A}_a$ and $y^A_a$ as the average values on an azimuth line that are used in $\hat{\phi}^A_a$, the term $\frac{x^A_a(i)-x^A_a(i+\rho(a))}{y^A_a(i)-y^A_a(i+\rho(a))}$ is
used to determine whether a sample's difference is dominated by the
change in intensity from the backscattering of the targets. We
determined an acceptable range as 0 to 2.5. The lower bound was
zero because a negative ratio
cannot be created from the additive noise, assuming the noise model is
correct. A liberal upper bound was chosen as 2.5 based on experimentation. Thus, the weighting term is
  defined as
  \begin{equation}
    w^{A}_a(i) = \left\{ \begin{array}{lr}
                          1 & \text{if }
                              0 < \frac{x^A_a(i)-x^A_a(i+\rho(a))}{y^A_a(i)-y^A_a(i+\rho(a))}  < 2.5\\
                          0 & \text{otherwise}
                        \end{array} \right\}
  \end{equation}
  to remove the outlying pairs from the loss function.
  
  As proven in the appendix, (\ref{eq:err_scallop}) can be represented using an inner product formulation
  \begin{equation}
    \begin{aligned}
      L^A = [\mathbf{v}^{A}-\mathbf{C}^{A}\mathbf{\hat{k}}]^T[\mathbf{v}^{A}-\mathbf{C}^{A}\mathbf{\hat{k}}].
  \end{aligned}
  \label{eq:err_scallop_mat}
  \end{equation}

  \subsection{Objective: Range}\hypertarget{tg:obj_rg}{}
  \label{sec:obj_rg}
  As noted previously, the noise pattern in the range direction
  is unique for each subswath. Thus, two objective function terms,
  $L^{R}$ and $L^{B}$, are proposed
  to compensate the noise within the subswaths (intra-subswath) and
  between the subswaths (inter-subswath).

  \subsubsection{Intra-subswath}
\begin{figure}[H]
  \begin{center}
    \includegraphics[width=0.5\linewidth]{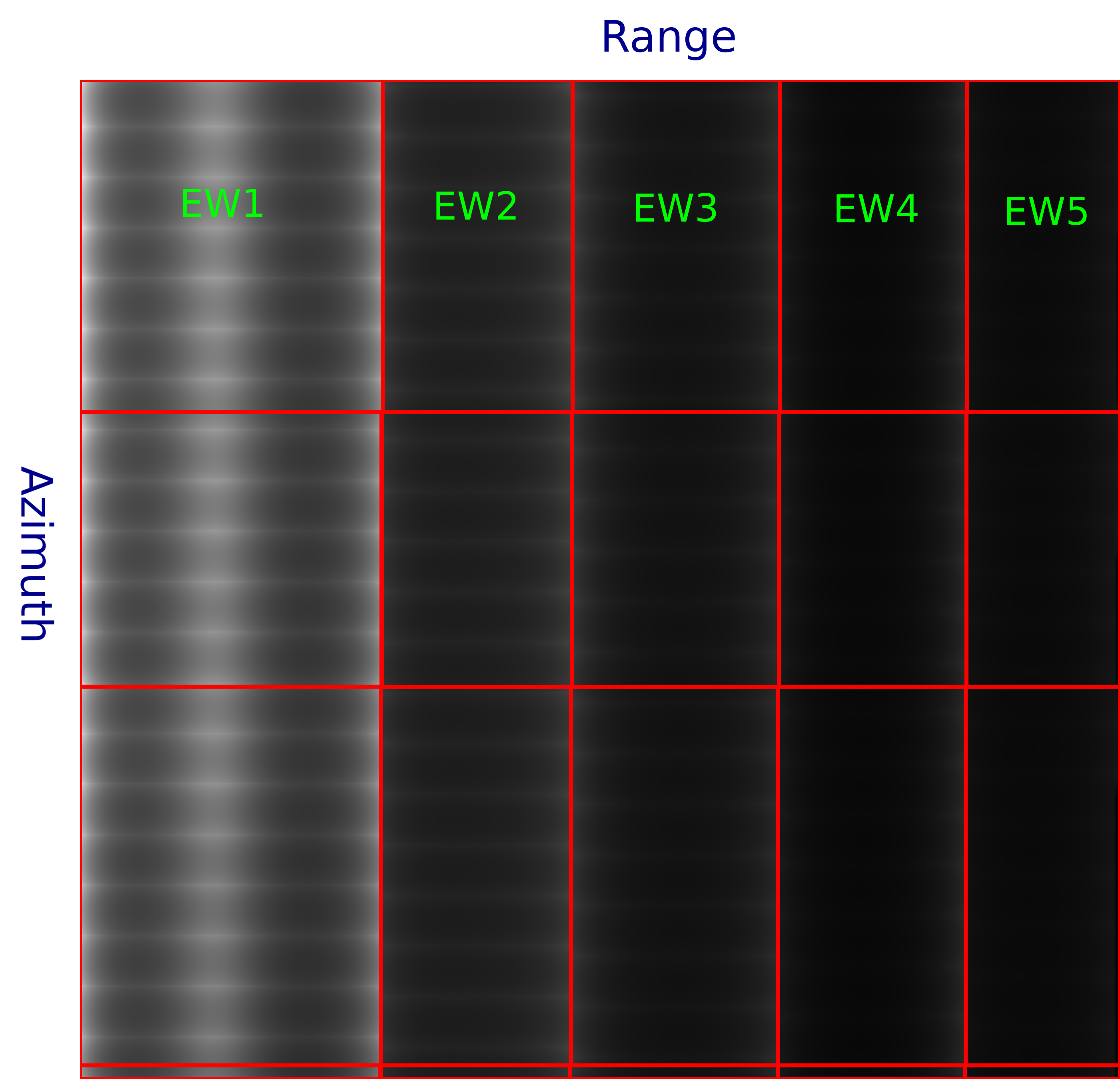}
    \end{center}
  \caption{Layout of the rectangular subregions $b$ in red overlaid with the estimated noise pattern for reference.}
  \label{fig:rectangle_layout}
\end{figure}

  First we examine the intra-subswath loss term that describes the noise pattern \textit{within} each
  subswath.  
Like the azimuth loss term, the intra-subswath loss term is constructed by taking the
squared difference between peaks and troughs but within the range component. The range noise pattern is neither periodic nor symmetric; instead the true scale of the noise is indirectly measured from taking the squared difference between peaks and
troughs exclusively. \figref{fig:rg_noise} shows the samples used with respect to the range direction. The samples used in the intra-subswath term are based on the
rectangular subregions divided along the azimuth of a subswath that
are dictated by the \textit{swathBoundList} field in the noise calibration XML file, as depicted in \figref{fig:rectangle_layout}. The intra-subswath
loss is based on the difference between adjacent peaks and troughs
over each of the subregions. Let $p_{b}(z)$ and $t_{b}(z)$ be
the $z^{\text{th}}$ peak and trough respectively within rectangular
subregion $b$ in subswath $a$. Now let $\hat{\phi}^R_b(s)$ be the mean
average of values predicted by the denoising model $\hat{\phi}$
within rectangular subregion $b$ in subswath $a$ that have a range
index within $s-\epsilon$ to $s+\epsilon$, with a
padding constant $\epsilon$ and $s$ being either $p_b(z)$ or $t_b(z)$. Since the range noise pattern is unimodal
in EW2, EW3, EW4, and EW5 (loss term $M$), while multi-modal in EW1 (loss term $N$), the intra-subswath
loss is composed as the sum of two terms
\begin{equation}
  \begin{aligned}
     L^{R} =& M + N,
   \end{aligned}
 \end{equation}
 where
\begin{equation}
  \begin{aligned}
    M = \sum_{a\in \mathcal{M}}\sum_{b \in a}  \big[&[w^R_b(1)[{\hat{\phi}}^R_b(p_b(1)) - {\hat{\phi}}^R_b(t_b(1))]]^2\\
    & + [w^R_b(2)[{\hat{\phi}}^R_b(t_b(1)) - {\hat{\phi}}^R_b(p_b(2))]]^2\big] \\
    &\text{such that } \mathcal{M} = \{\text{EW2, EW3, EW4, EW5}\}
  \end{aligned}
\label{eq:M}
\end{equation}
and
\begin{equation}
  \begin{aligned}
    N = \sum_{b \in \text{EW1}} \big[ &[w^R_b(1)[{\hat{\phi}}^R_b(p_b(1)) - {\hat{\phi}}^R_b(t_b(1))]]^2 \\
    &+[w^R_b(2)[{\hat{\phi}}^R_b(t_b(1)) - {\hat{\phi}}^R_b(p_b(2))]]^2 \\
    & + [w^R_b(3)[{\hat{\phi}}^R_b(p_b(2)) - {\hat{\phi}}^R_b(t_b(2))]]^2 \\
    & + [w^R_b(4)[{\hat{\phi}}^R_b(t_b(2)) - {\hat{\phi}}^R_b(p_b(3))]]^2 \big]
  \end{aligned}
  \label{eq:N}
\end{equation}

As in the azimuth noise component, weighting terms $w^R$ were
introduced to account for instances where the difference is dominated
by the difference in land targets (e.g. land vs
ice). In this case the weighting term is defined as
\begin{equation}
  w^{R}_b(o) = \left\{ \begin{array}{lr}
                         \mu & \text{if }
                             0 < \frac{{x}^{R1}_b-{x}_b^{R2}}{{y}_b^{R1}-{y}_b^{R2}}  < 2.5\\
                         0 & \text{otherwise}
                       \end{array} \right\},
                   \end{equation}
                   where $\hat{\phi}^{R}_b(p_b(o+q)) - \hat{\phi}^{R2}_b(t_b(o)) = [x^{R1}_b
                   - k_ay^{R1}_b] -  [x^{R2} - k_ay^{R2}_b]$ with $o
                   \in \{1,2\}$ and $q \in \{0,1\}$ such that they
                   correspond with the difference terms in (\ref{eq:M}) and (\ref{eq:N}). The
                   value $\mu$ balances the trade-off between
                   $L^A$, which is summed over a larger number of
                   terms. From experimentation we chose
                   $\mu = 1.79$.

Finally, the proposed error for the intra-subswath range noise can be
rewritten as an inner product
\begin{equation}
  \begin{aligned}
     L^{R} &= [\mathbf{v}^{R}-\mathbf{C}^{R}\mathbf{\hat{k}}]^T[\mathbf{v}^{R}-\mathbf{C}^{R}\mathbf{\hat{k}}],
  \end{aligned}
  \label{eq:range_loss_mat}
\end{equation}
as shown in the appendix.

\subsubsection{Inter-subswath}
As the radiation pattern changes when the antenna switches between
subswaths, we must consider that this will cause discontinuities between
adjacent subswaths. Intuitively, the range columns of adjacent subswaths are spatially
  correlated and should have similar values in an ideal denoised
  image. We introduce another loss term to
  represent the difference of intensity between the columns
  of adjacent subswaths to handle these considerations.
  
Let $\hat{\phi}_b^{B1}$ be the average predicted value of the
last $\epsilon$ range columns that are within subrectangle $b$ in
subswath $a$ and let $\hat{\phi}_b^{B2}$ be the average of the
$\epsilon$ range columns after the last column in subrectangle $b$. Then the inter-subswath loss term is determined by 
    \begin{equation}
    \begin{aligned}
      L^{B} &= \sum_{a \in \mathcal{B}} \sum_{b \in a} [\hat{\phi}_b^{B1} -
      \hat{\phi}_b^{B2}]^2\\
      &= [\mathbf{v}^{B}-\mathbf{C}^{B}\mathbf{\hat{k}}]^T[\mathbf{v}^{B}-\mathbf{C}^{B}\mathbf{\hat{k}}] \\
      & \text{such that } \mathcal{B} = \{\text{EW1, EW2, EW3, EW4}\}
  \end{aligned}
    \label{eq:interswath_err}
  \end{equation}

  \subsection{Regularization}
  \label{sec:reg}
  Even with weight terms in the loss function, using each of the
  aforementioned loss terms can result in unstable estimates
  $\mathbf{\hat{k}}$. To resolve this, a prior assumption is made that each
  value of $\mathbf{\hat{k}}$ should be close to 1. Thus a regularization
  term is introduced to penalize estimates of $\mathbf{\hat{k}}$ that diverge from 1
  \begin{equation}
    \begin{aligned}
      L^{r} &= \sum_a [\lambda_a[1-\hat{k}_a]]^2 \\
      &= [\mathbf{1} - \mathbf{C}^r\mathbf{\hat{k}}]^T[\mathbf{1} - \mathbf{C}^r\mathbf{\hat{k}}],
    \end{aligned}
  \end{equation}
  where $\mathbf{C}^r$ is the diagonal matrix containing the hyper-parameter
  vector $\bm{\lambda}$. By applying the Golden section search algorithm in a block-wise manner, we found appropriate values of $\bm{\lambda} = [0.1,0.1,6.75124,2.78253,10]$, used for all images. Now that we have introduced the loss terms, we now detail
  the implementation for estimating the scaling parameters for each subswath.
  
  \subsection{Implementation}
  \label{sec:obj_impl}

  Recall that the standard denoising method proposed by ESA recommends denoising as
  \begin{equation}
    \phi_a = x_a - y_a.
  \end{equation}
  As it is apparent that this is not sufficient in some scenarios,
the desired model scales the noise factor according to each subswath $a$
with $\hat{k}_a$
  \begin{equation}
    \hat{\phi}_a = x_a  - \hat{k}_ay_a
    \label{eq:scale_denoise}
  \end{equation}
Through matrix and vector concatenation the final loss term is defined
as the sum of all the previously derived terms in a single inner product
\begin{equation}
  \begin{aligned}
  L &= L^{A} + L^{R} + L^{B} + L^{r}\\
  &= [\mathbf{v}-\mathbf{C}\mathbf{\hat{k}}]^T[\mathbf{v}-\mathbf{C}\mathbf{\hat{k}}]
  \end{aligned}
\end{equation}
and the solution for $\underset{\mathbf{\hat{k}}}{\mathrm{argmin}}\ L$ is
well known \citep{cambridge_convex,fieguth} to be the least-squares solution
\begin{equation}
  \mathbf{\hat{k}} = [\mathbf{C}^T\mathbf{C}]^{-1}\mathbf{C}^T\mathbf{v}.
  \label{eq:master}
\end{equation}

\section{Experiments}
\label{sec:experiments}
To evaluate the effectiveness of our method, two experiments
were considered. The first experiment is a simulation where a set of
RADARSAT SAR images without significant additive noise patterns are selected to demonstrate the
abilities of the proposed method in a setting where the characteristics of
the noise are directly controlled. The second experiment considers denoising Sentinel-1 EW
images in both HV and VH polarization and evaluates the effectiveness by examining sub-regions of open-water.

\subsection{Parameter re-estimation simulation on RADARSAT}
\label{sec:sent3}

The simulation experiment can be summarized by the following: first
RADARSAT images were selected, a scaled noise
field was added to the images, scaling parameters were re-estimated on
the noisy image, followed by denoising the noisy image with
the model and comparing the result with original ground truth images. The RADARSAT images were selected to ensure that a ground
truth reference image was available and independent of the additive
noise patterns specific to Sentinel-1 EW. 
The goal of this experiment is to test if the proposed method
can accurately estimate the optimal scaling parameters and verify the
quality of the denoised image compared to a true original.

The set of images used for the experiments were 20 RADARSAT HV images taken in the Beaufort Sea during 2010. A template SAR noise field was selected from a sample Sentinel-1 scene\footnote{The sample scene was from S1A\textunderscore
EW\textunderscore GRDM\textunderscore 1SDH\textunderscore
20180902T164932\textunderscore 20180902T165032\textunderscore
023522\textunderscore 028FAA\textunderscore 5A8B}. To ensure comparability, the values of the
RADARSAT scene prior to adding noise were re-scaled to have a comparable signal to noise
ratio as the original Sentinel-1 scene after denoising with the proposed method. The RADARSAT scene was also spatially re-scaled with linear interpolation to have the same number of rows and
columns as the Sentinel-1 scene.

\begin{figure}[H]
\centering
  \begin{tabular}{|M{34mm}M{34mm}M{34mm}M{34mm}|}\hline
    \scriptsize{Original} & \scriptsize{Original+Noise}  & \scriptsize{Baseline} & \scriptsize{Proposed}  \\\hline
     \includegraphics[width=\linewidth]{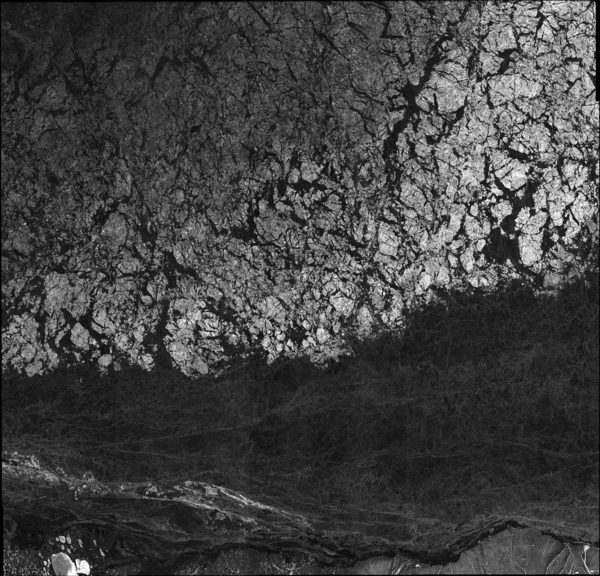} & \includegraphics[width=\linewidth]{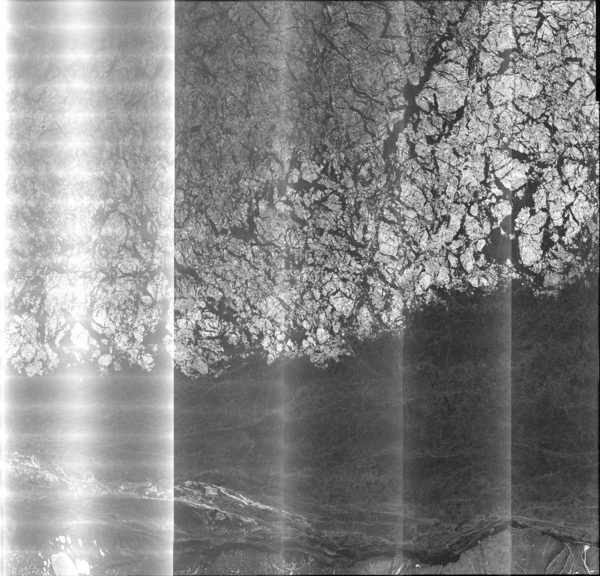} &\includegraphics[width=\linewidth]{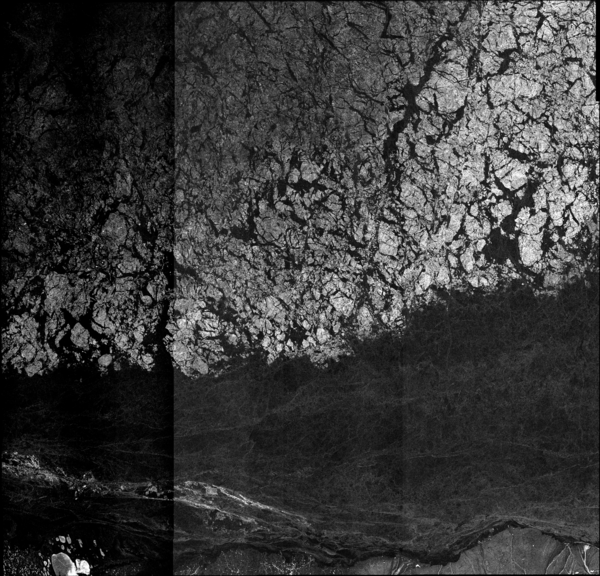} & \includegraphics[width=\linewidth]{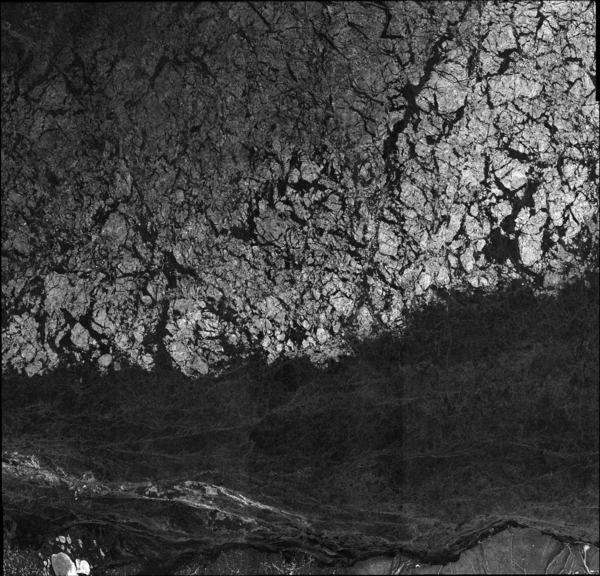}   \\
   \includegraphics[width=\linewidth]{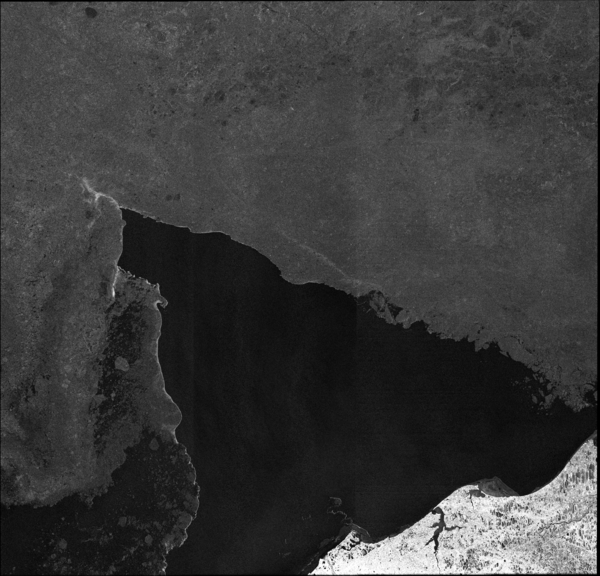} & \includegraphics[width=\linewidth]{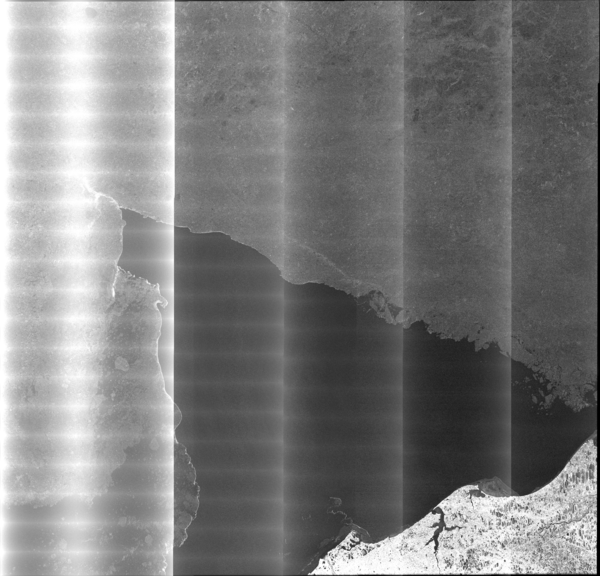} &\includegraphics[width=\linewidth]{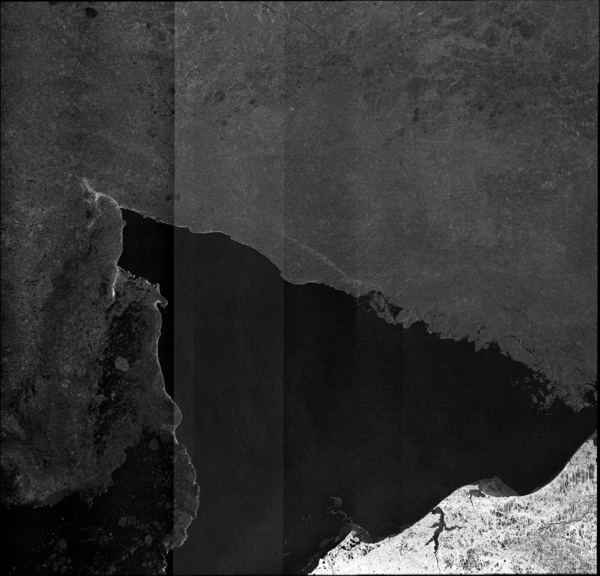} & \includegraphics[width=\linewidth]{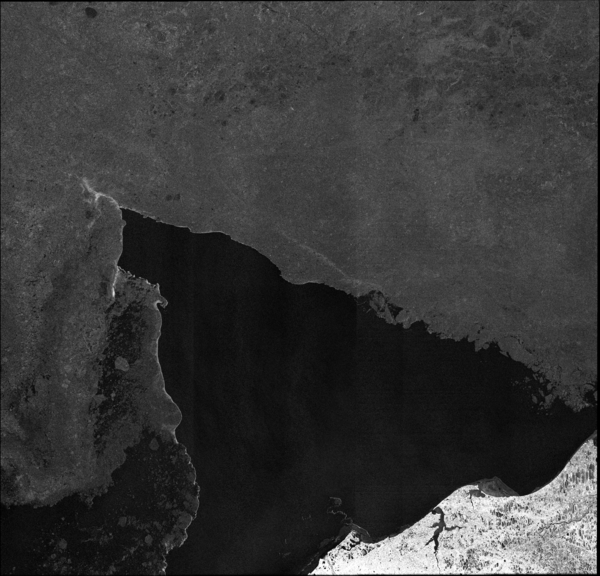}    \\
    \includegraphics[width=\linewidth]{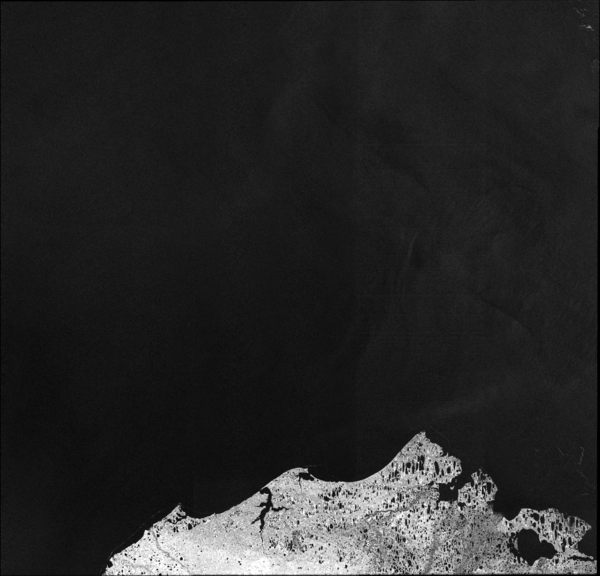} & \includegraphics[width=\linewidth]{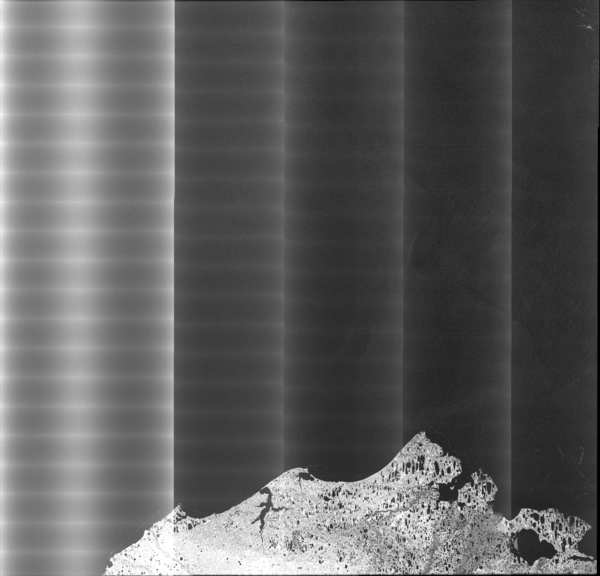} &\includegraphics[width=\linewidth]{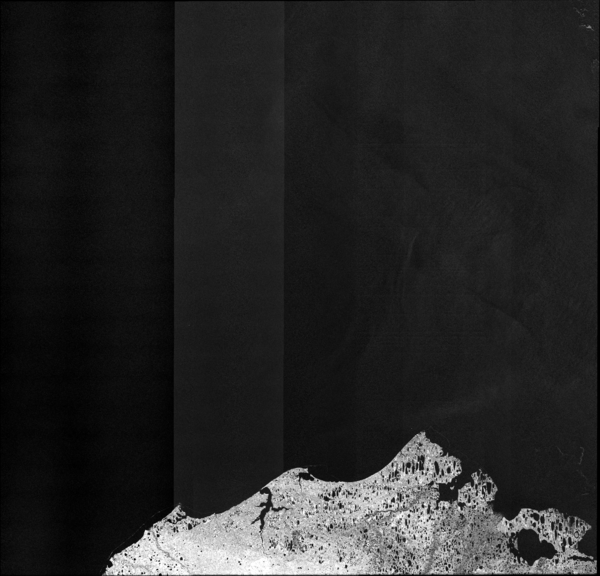} & \includegraphics[width=\linewidth]{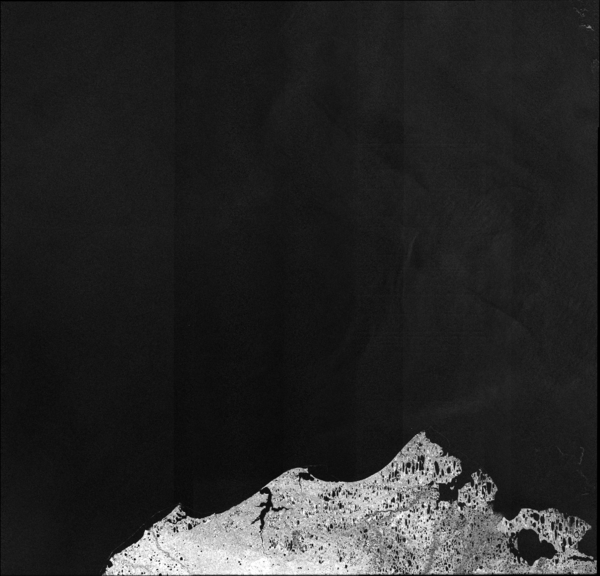}  \\\hline
  \end{tabular}
    \caption{Comparison of images within the simulation experiment with $\hat{k} = [1.55, 0.833, 1.01,  1.01, 1.01]$.
The proposed method is almost identical to the original image, while the
image from the baseline method is scaled improperly between the
simulated first and second subswaths.}
\label{tab:sim_pictures}
  \end{figure}

For each image, 10 different noisy versions were constructed with scaling factors randomly
selected from uniform distributions within the ranges $k_{\text{EW1}}: [1.2, 1.6],  k_{\text{EW2}}: [0.8,1.0],
k_{\text{EW3}}: [0.92,1.02], k_{\text{EW4}}:[0.95,1.05], k_{\text{EW5}}:
[0.98,1.02]$, resulting in a total of 110 noisy images. These ranges for $\mathbf{k}$ were selected as they were representative of the
range of estimated parameters encountered in the sample of true SAR
images discussed in the following subsection. The noisy images were created by multiplying the noise field
with the scaling parameters and adding it with the original image. Then, using (\ref{eq:master}),
the parameters were estimated, with the scaled subtraction labelled
as the \textbf{proposed} denoising. As a \textit{baseline}, we perform the
aforementioned procedure using scaling estimates as
$\bar{\mathbf{k}} = [1.4, 0.925, 0.985, 1.0, 1.0]$, which are the
central values of the distributions. In addition, a visual comparison is shown in
\figref{tab:sim_pictures}.

Three metrics, the normalized root mean squared error (NRMSE), peak signal to noise ratio (PSNR), and structural similarity
index (SSIM) \citep{ssim}, were
computed with respect to the original image to determine the
effectiveness of denoising. NRMSE computes the
square root of the aggregate sum of pixel-wise
squared differences, with a lower-bound of 0 indicating a perfect
comparison. In contrast to root mean squared error, NRMSE is
normalized by dividing the root sum of squares by the range of input values for
each image, which helps ground the metric against a comparison that vary
in scale \citep{nrmse}. We use a common variant of NRMSE as
\begin{equation}
  NRMSE(\mathbf{d}, \mathbf{l}) = \frac{\sqrt{MSE(\mathbf{d},\mathbf{l})}}{max(\mathbf{l})-min(\mathbf{l})},
\end{equation}
where $MSE(\mathbf{s},\mathbf{l})$ is the mean squared error between prediction vector $\mathbf{d}$ and baseline vector $\mathbf{l}$. PSNR computes the logarithmic ratio of power of the original image to the power of the noise
within the modified image. PSNR is commonly used in applications that
result in distortion of the original image in denoising or compression
applications \citep{gonzalez, psnr_video}. Finally, SSIM is a metric bounded between 0 and 1,
with 1 indicating the two images are identical, and is commonly used
for evaluating denoising and reconstruction methods by quantifying the similarity of
structural features within an image \citep{ssim}. Each of these metrics make
use of the available ground truth reference image. The summary of the metrics with respect to the denoising procedures over the distribution of images are shown in \tabref{tab:param_stats}, with the metrics for the applying no denoising (\textit{noisy}) included for context.

To compare the generated distributions between the proposed and baseline methods for each of the metrics, we employed a one-tailed paired
t-tests using a critical value $\alpha=0.05$. We found our method had significantly higher SSIM and
PSNR and lower NRMSE than the baseline method ($p < \alpha$), indicating it is significantly better
in terms of all three metrics.

\begin{table}[H]
    \caption{Mean $\pm$ standard deviation for the normalized root mean squared
    error (NRMSE), peak signal to noise ratio (PSNR), and
    structural similarity index (SSIM) over all the
    images with 10 iterations of sampling values for $k$. P-values from the one-tailed t-test comparing the \textbf{proposed} method to the other methods are also shown.  }
\label{tab:param_stats}
  \begin{center} \resizebox{\textwidth}{!}{
  \begin{tabular}{|c||cc|cc|cc|}\hline
    Method & NRMSE & p-value & PSNR & p-value &  SSIM & p-value \\\hline
    \textit{Noisy} &  0.514 $\pm$ 0.028 & $\approx 0.0$ & 20.9 $\pm$ 2.0 & $\approx 0.0$ & 0.913 $\pm$ 0.009 & $\approx 0.0$\\
    \textit{Baseline} $\bar{\mathbf{k}}$ & 0.061 $\pm$  0.031 & $1.5 \times 10^{-47}$ & 41.4  $\pm$ 5.6 & $3.8 \times 10^{-57}$  & 0.993  $\pm$ 0.005 & $9.8 \times 10^{-23}$ \\
    \textbf{Proposed} & \textbf{0.017} $\pm$ \textbf{0.008} & N/A & \textbf{ 51.4} $\pm$ \textbf{3.8} & N/A &  \textbf{0.997} $\pm$ \textbf{0.001} & N/A \\\hline
  \end{tabular}}
  \end{center}
\end{table}

\subsection{Denoising Sentinel-1 SAR}
\label{sec:exp_line}
For the second experiment, a set of 41 Sentinel-1 GRDM SAR images was selected
to include 13 images from the Arctic Ocean, 6 from the Antarctic Ocean, 8 from the
Atlantic Ocean, 6 from the Pacific Ocean, and 8 from the Indian
Ocean. The images were selected from all five oceans to ensure our evaluation would not be biased towards a single environment. The images from the Arctic and Antarctic oceans have HV polarization while the
images from the Atlantic, Pacific, and Indian oceans have VH
polarization. The polarization of the images is determined by the Sentinel-1
observational scenario \citep{sent1_scenario}. In the remainder of this section we provide
both a visual and a quantitative comparison of the proposed and
baseline denoising procedures.

\begin{figure}
  \begin{minipage}[t]{\linewidth}
    \vspace{0pt}
    \begin{tabular}{|p{1cm}|M{44mm}M{44mm}M{44mm}|}\hline
      \scriptsize{Ocean div.}  & \scriptsize{Original} &
                                                         \scriptsize{Unscaled (ESA)} & \scriptsize{Proposed} \\\hline
    \scriptsize{Arctic}\newline\tiny{(73.5$^{\circ}$N, 141.2$^{\circ}$W)} &
             \includegraphics[width=\linewidth]{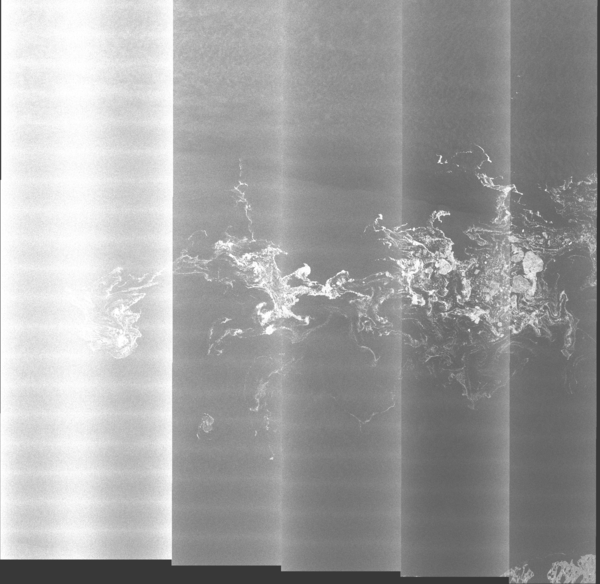}
               &
                 \includegraphics[width=\linewidth]{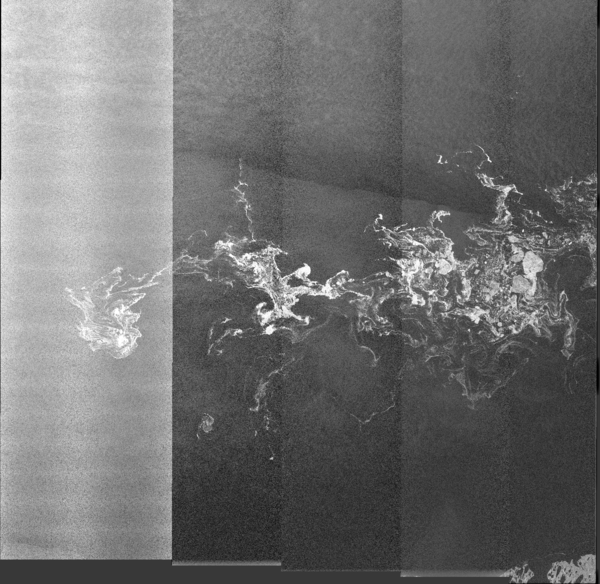}
                          &
                            \includegraphics[width=\linewidth]{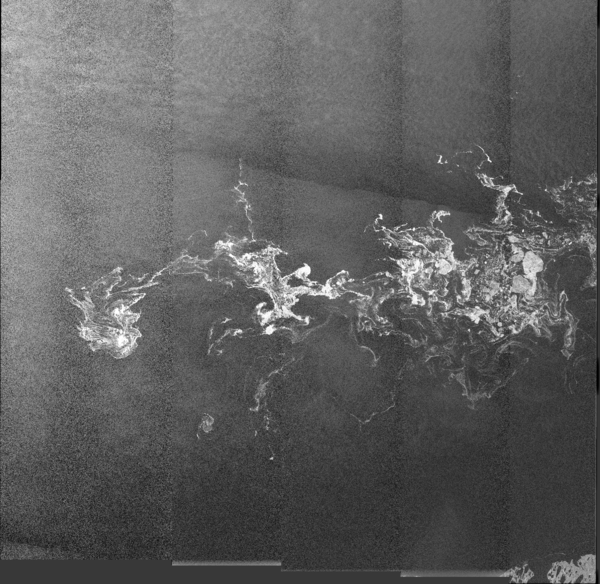}\\
    \scriptsize{Arctic}\newline\tiny{(73.6$^{\circ}$N, 136.9$^{\circ}$W)} &
             \includegraphics[width=\linewidth]{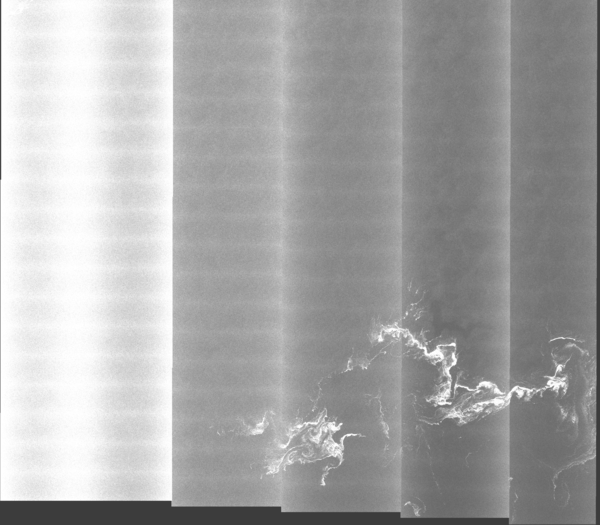}
               &
                 \includegraphics[width=\linewidth]{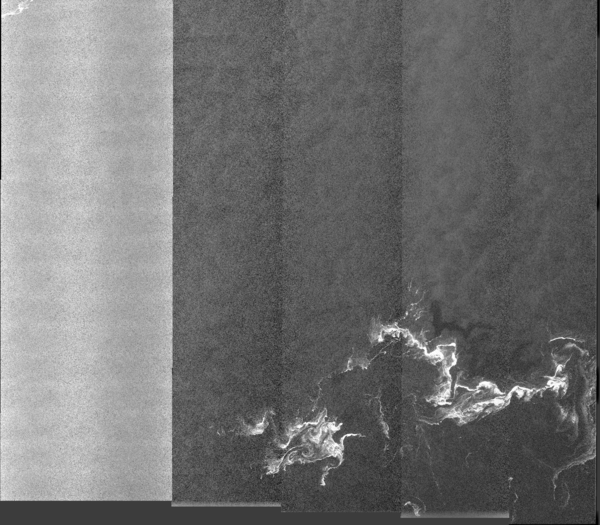}
                          &
                            \includegraphics[width=\linewidth]{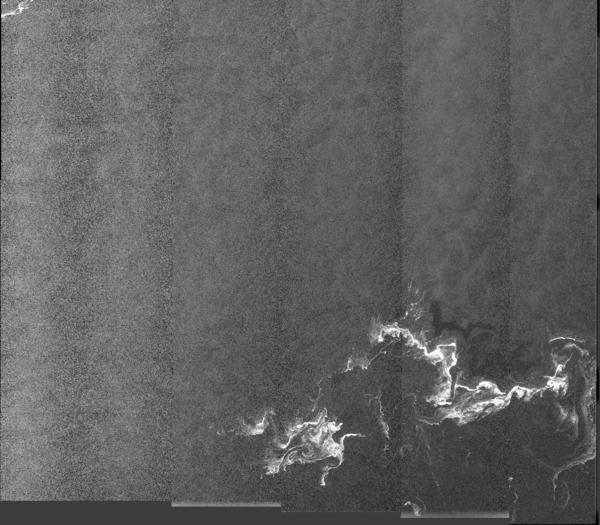}\\
    \scriptsize{Antarctic}\newline\tiny{(65.4$^{\circ}$S, 138.7$^{\circ}$W)} &
                \includegraphics[width=\linewidth]{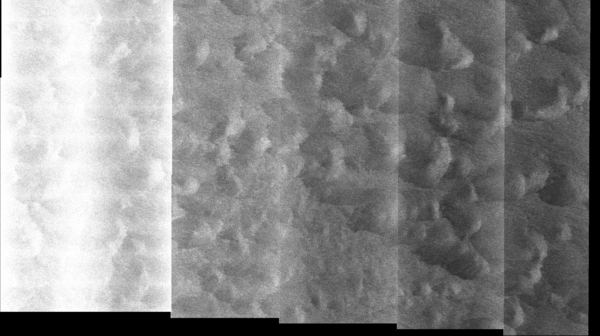}
               &
                 \includegraphics[width=\linewidth]{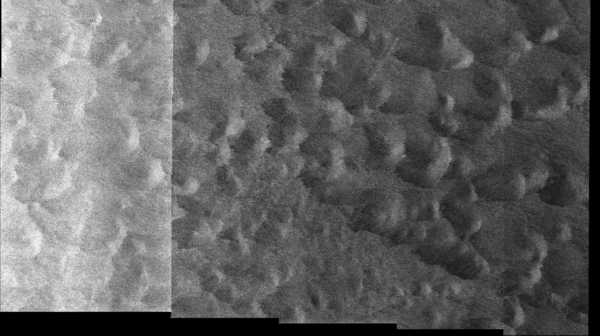}
                 &
                   \includegraphics[width=\linewidth]{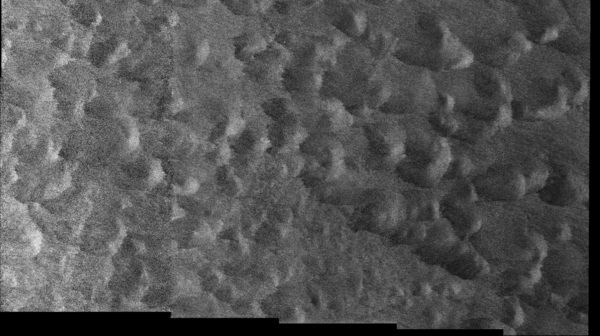}\\
   \scriptsize{Antarctic}\newline\tiny{(71.8$^{\circ}$S, 128.7$^{\circ}$W)}&
                \includegraphics[width=\linewidth]{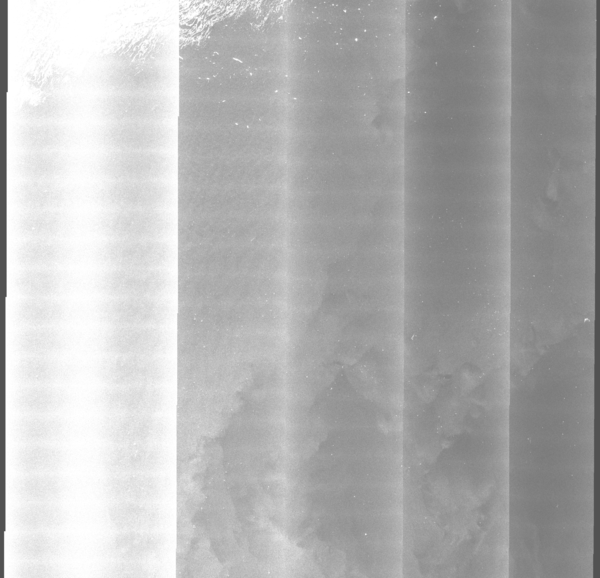}
               &
                 \includegraphics[width=\linewidth]{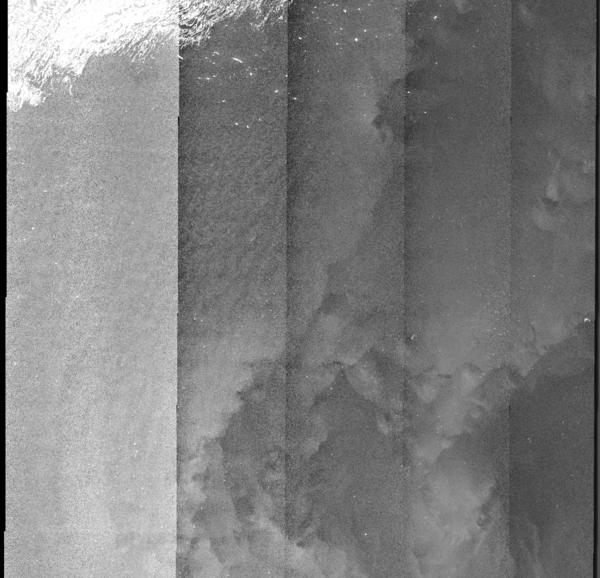}
                          &
                            \includegraphics[width=\linewidth]{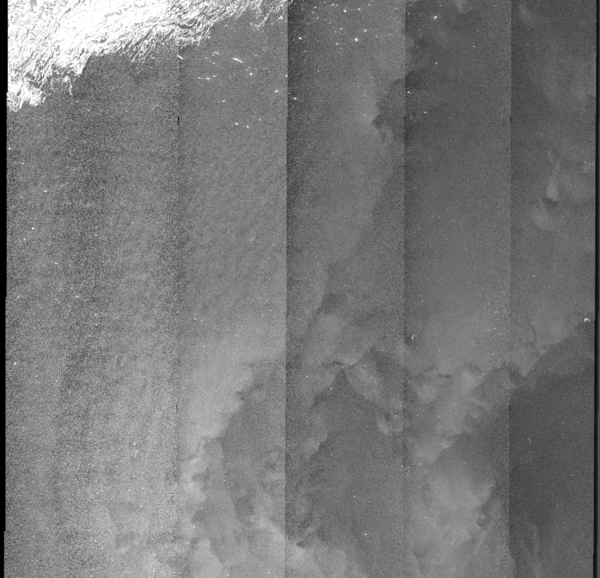}\\
    \scriptsize{Atlantic}\newline\tiny{(41.4$^{\circ}$N, 28.5$^{\circ}$W)} &
               \includegraphics[width=\linewidth]{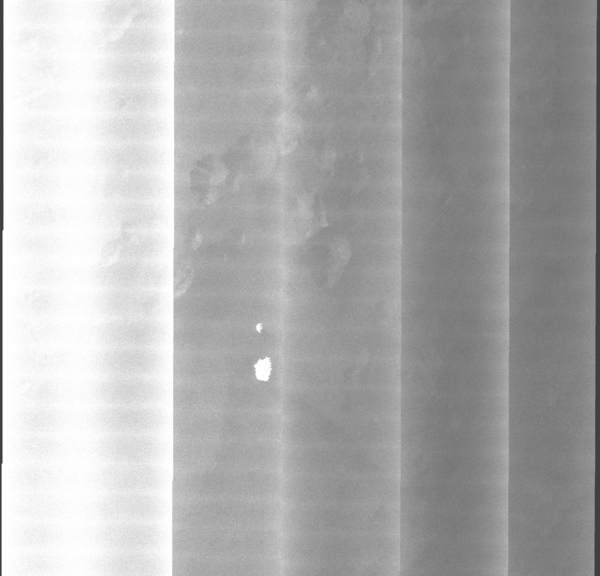}
               &
                 \includegraphics[width=\linewidth]{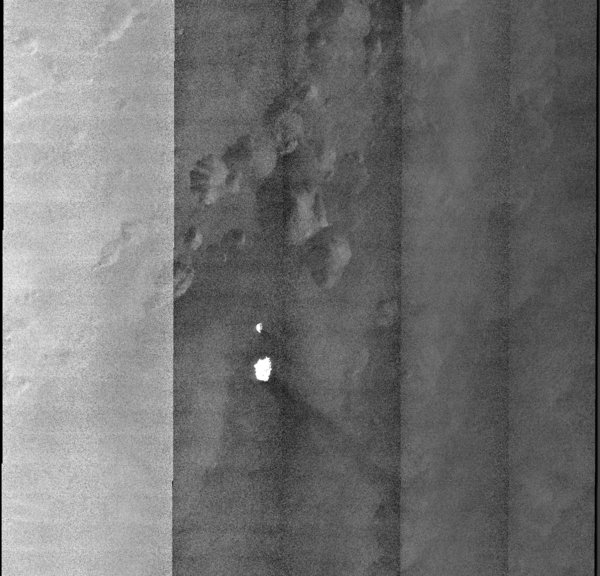}
                                    &
                                      \includegraphics[width=\linewidth]{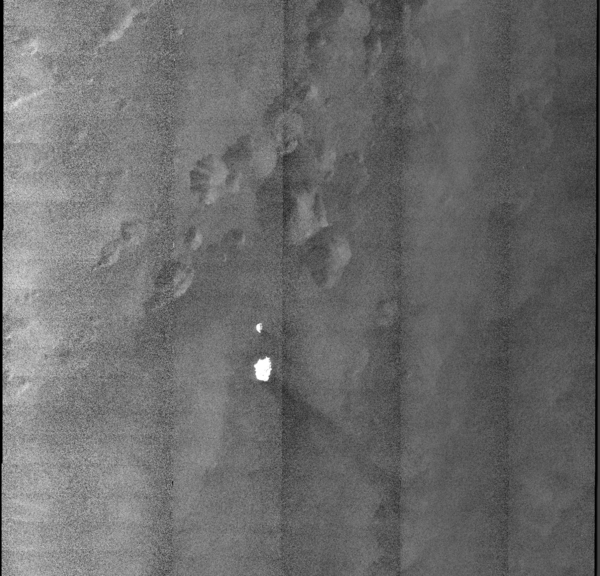}\\\hline
    \end{tabular}
  \end{minipage}
      \caption{Part 1 gallery of SAR images showing the effect of the baseline and the proposed noise removal methods in reference to the original image.}\label{fig:gal1}
  \end{figure}

\begin{figure}
   \begin{minipage}[t]{.5\linewidth}
        \begin{tabular}{|p{1cm}|M{44mm}M{44mm}M{44mm}|}\hline
      \scriptsize{Ocean div.}  & \scriptsize{Original} &
                                                         \scriptsize{Unscaled (ESA)} & \scriptsize{Proposed} \\\hline
    \scriptsize{Atlantic}\newline\tiny{(37.8$^{\circ}$N, 26.8$^{\circ}$W)} &
               \includegraphics[width=\linewidth]{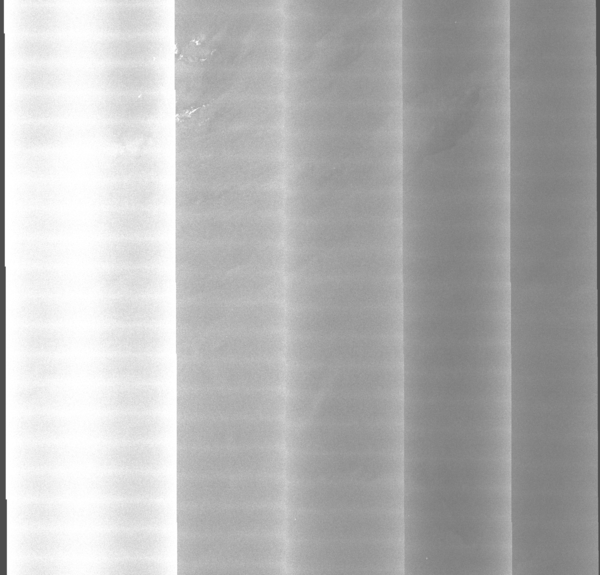}
               &
                 \includegraphics[width=\linewidth]{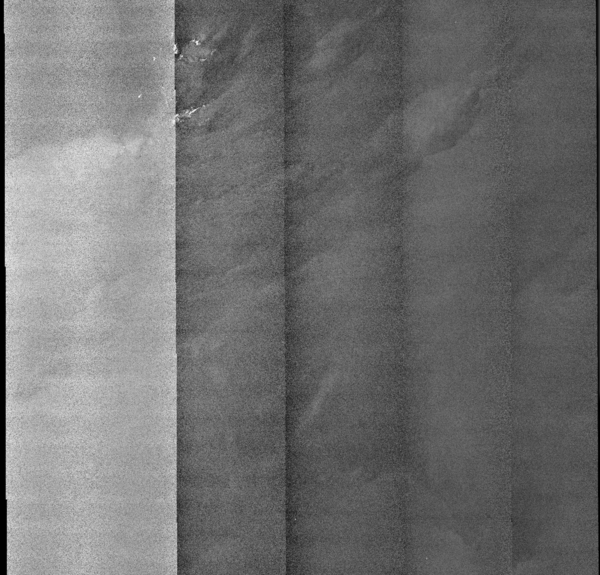}
                          &
                            \includegraphics[width=\linewidth]{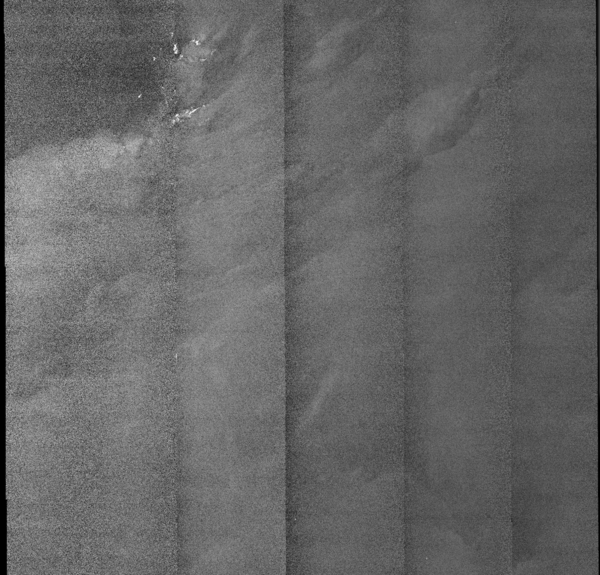}\\
    \scriptsize{Pacific}\newline\tiny{(16.5$^{\circ}$N,  150.9$^{\circ}$E)} &   \includegraphics[width=\linewidth]{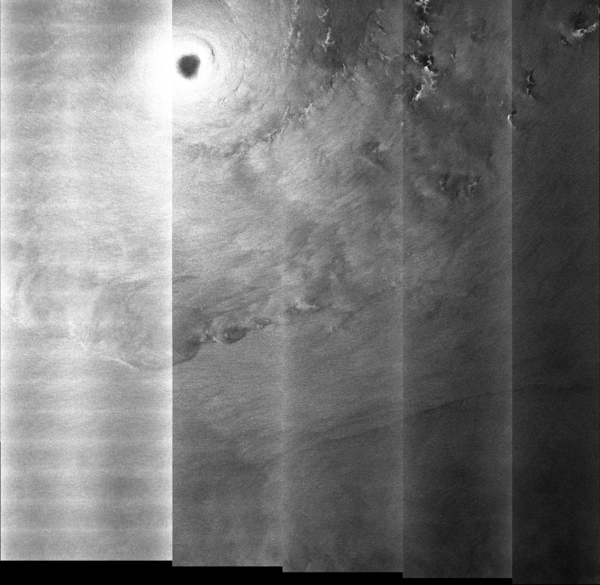}
               &
                 \includegraphics[width=\linewidth]{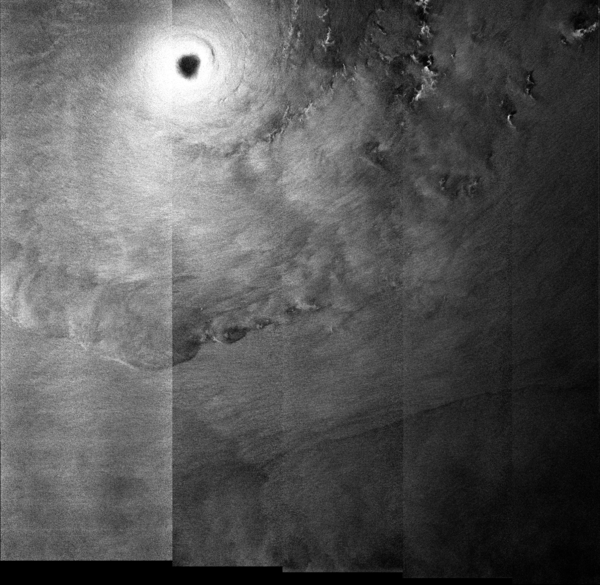} &
                                                                                          \includegraphics[width=\linewidth]{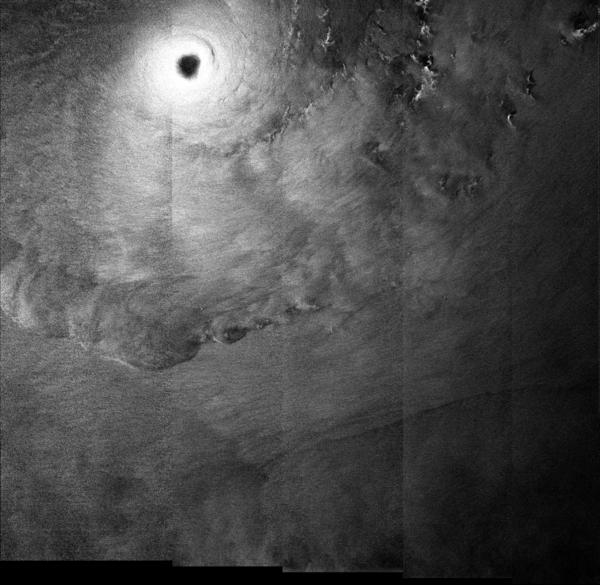}\\
    \scriptsize{Pacific}\newline\tiny{(19.7$^{\circ}$N, 174.7$^{\circ}$W)} &\includegraphics[width=\linewidth]{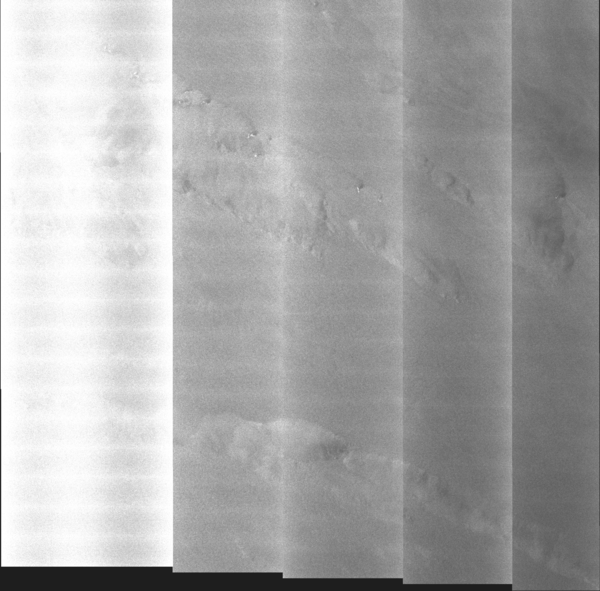}
               &
                 \includegraphics[width=\linewidth]{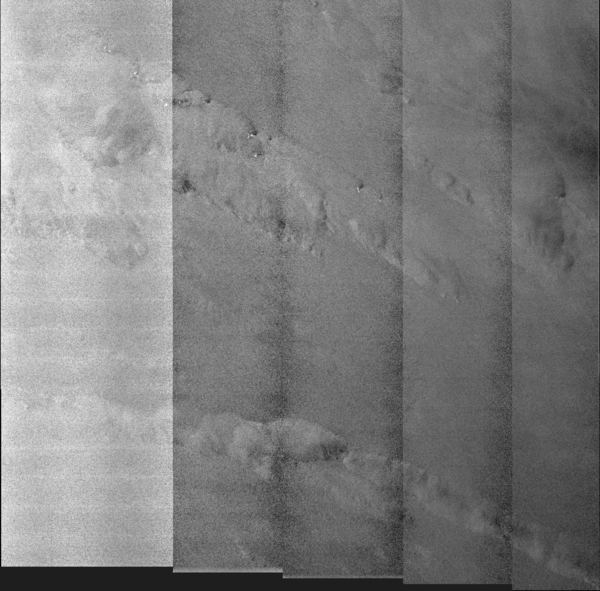} &
                                                                                          \includegraphics[width=\linewidth]{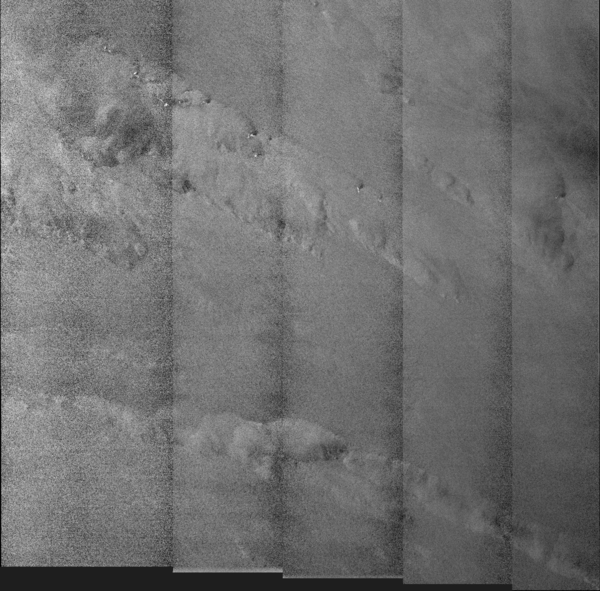}\\
    \scriptsize{Indian}\newline\tiny{(21.1$^{\circ}$S, 74.8$^{\circ}$E)}&
             \includegraphics[width=\linewidth]{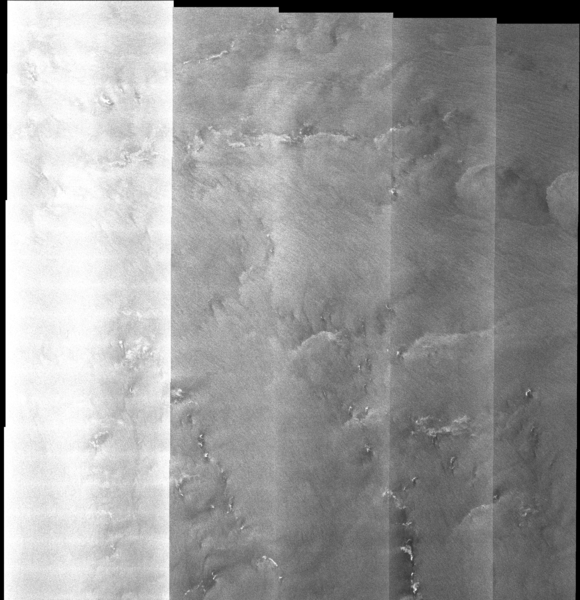}
               &
                 \includegraphics[width=\linewidth]{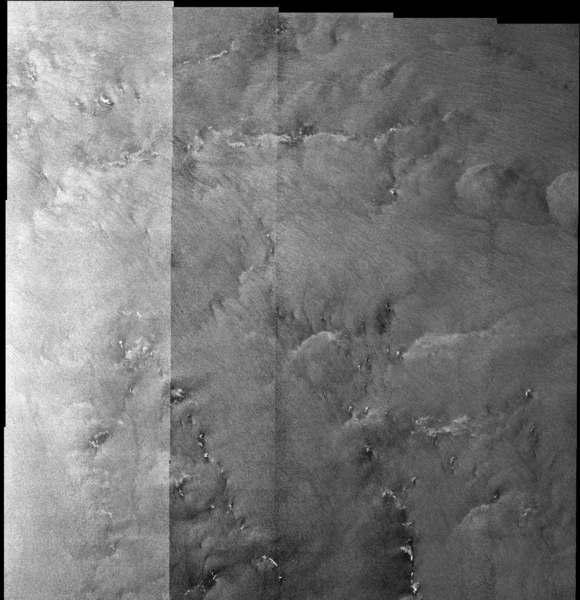}
                          &
                            \includegraphics[width=\linewidth]{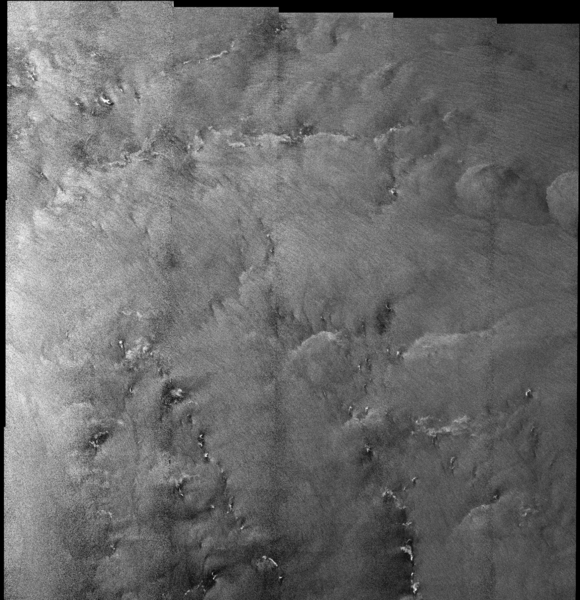}\\
    \scriptsize{Indian}\newline\tiny{(21.3$^{\circ}$S, 76.9$^{\circ}$E)}&
             \includegraphics[width=\linewidth]{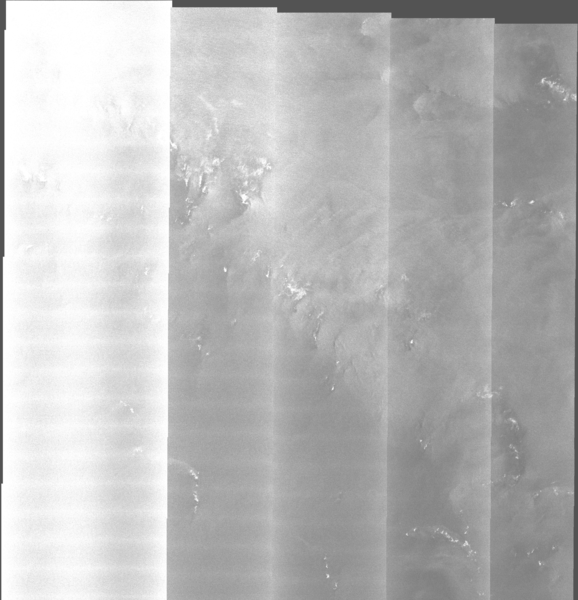}
               &
                 \includegraphics[width=\linewidth]{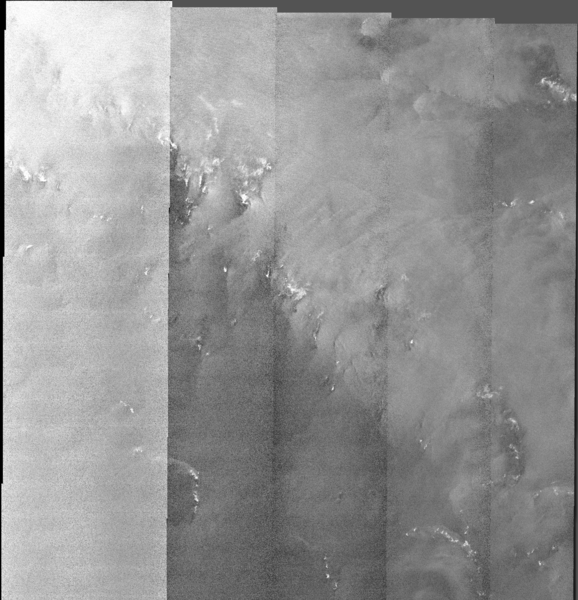}
                          &
                            \includegraphics[width=\linewidth]{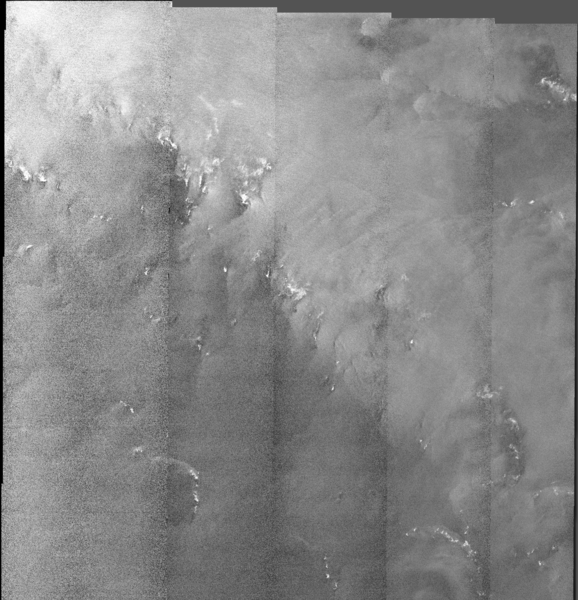}\\\hline
  \end{tabular}
\end{minipage}
      \caption{Part 2 gallery of SAR images showing the effect of the baseline and the proposed noise removal methods in reference to the original image.}\label{fig:gal2}
\end{figure}

\figref{fig:gal1} and \figref{fig:gal2} visually present a subset of co-polarized images from all five major ocean divisions. Three types
of images are shown: the original image, the scene with ESA denoising, and the scene with a scaled noise field scaled and
subtracted via the proposed method. For display purposes only, the images were linearly
scaled from 0 to 255 for each scene based on the 2.5 to 97.5 percentile of the
distribution of intensities for all versions of the
scene in order to maximize the contrast in each image.
Visually, the proposed method greatly reduces the noise characteristics
from each of the images, although there are still some discontinuities in
some of the images. This is not an error in the parameter estimation,
but rather in the assumption that the noise field is linearly mis-scaled. More detail regarding this is provided in \secref{sec:discussion}.

\begin{figure}[H]
  \begin{subfigure}{\linewidth}
    \includegraphics[width=\linewidth]{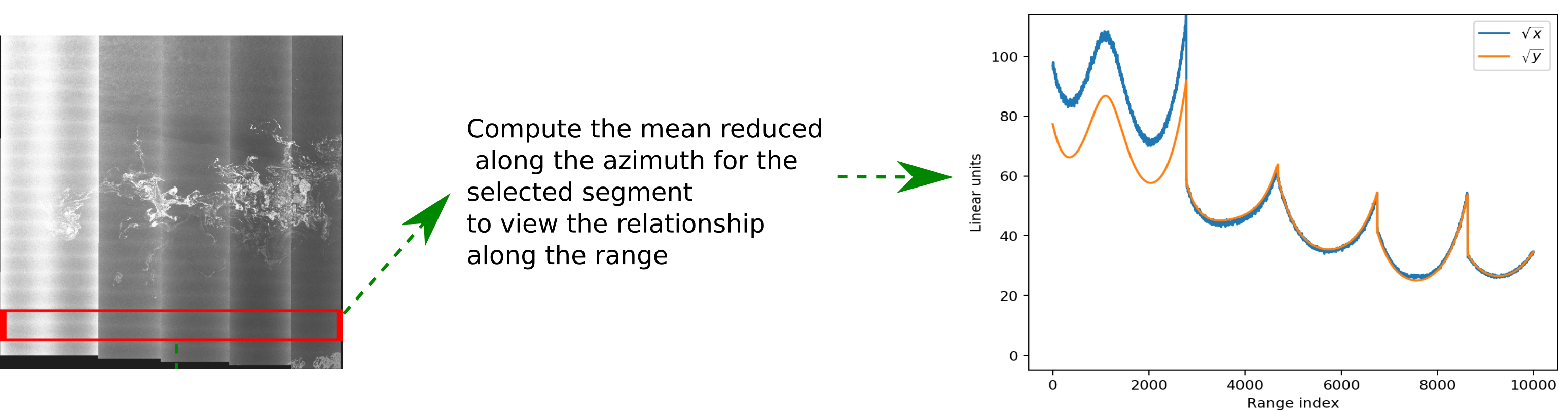}
  \caption{Ocean value selection process}
  \label{fig:exp_explain}
\end{subfigure}
\begin{subfigure}{\linewidth}
  \begin{center}
    \includegraphics[width=0.7\linewidth]{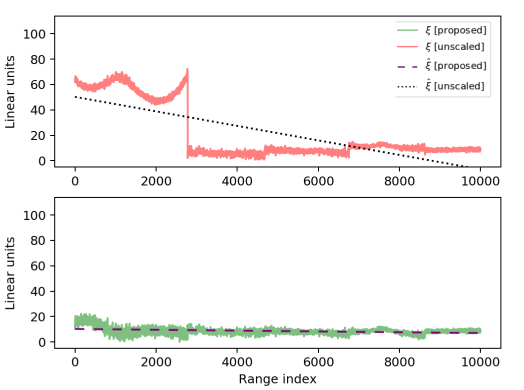}
    \end{center}
  \caption{Corresponding linear regression for the ESA method (top) 
    and our method
    (bottom) using the mean values selected from \figref{fig:exp_explain}.}
\end{subfigure}
\caption{Visualization of the quantitative experiment detailed in
  section \ref{sec:exp_line}}
\label{fig:exp_line}
\end{figure}

For the quantitative comparison of the experiment, we selected rectangular sub-regions in
the image that spanned the entire range (all five subswaths) and were over open-water
without major textural features. We make the assumption that the
intensity over such sub-regions will not vary significantly with
respect to range when the image is ideally denoised. A similar method of evaluation was also used in \cite{thermal}. Consequently, we
assume that the mean intensity of these sub-regions should follow a
linear relationship, with deviation from a linear relationship implying the presence of noise.
Let $\xi(j)$ be the mean value along the azimuth on range index $j$
for a denoised image. Let $\hat{\xi}(j)$ be the linear fit for
$\xi(j)$, which as previously mentioned, should be a reasonable estimate for $\xi(j)$. \figref{fig:exp_line} shows an
example of this process. The NRMSE between $\xi(j)$ and $\hat{\xi}(j)$ was
then compared for the ESA denoising method and our proposed
denoising method.
The mean and standard deviation of NRMSE over all images are shown in
\tabref{tab:NRMSE_results}. Using a one-tailed paired t-test, we found that our method had
significantly lower NRMSE than 
the ESA method ($p = 9.8
\times 10^{-11} < 0.05$). 

\begin{table}[H]
  \caption{Normalized root mean squared error for each image over
    selected ocean sub-images along with p-values generated from a
    one-tailed T-test comparing the proposed method to the others. The table is divided into comparing data from IPF 2.91 specific to sections \ref{sec:exp_line} and \ref{sec:static}, and 3+ that is specific to section \ref{sec:ipf3}}
  \label{tab:NRMSE_results}
  \begin{center}
    \begin{tabular}{|c|cc||cc|}\hline
      & \multicolumn{2}{c||}{IPF 2.91} & \multicolumn{2}{c|}{IPF 3+}\\
    Method & NRMSE & p-value & NRMSE & p-value \\\hline
    \textit{ESA} & 1.85 $\pm$ 0.941 & $9.8 \times 10^{-11}$ & 0.685 $\pm$ 0.278 & 0.017 \\
    \textit{Static} $\bar{k}$ & 0.814 $\pm$ 0.674 & $0.0058$ & N/A & N/A  \\
    \textbf{Proposed} & \textbf{0.568} $\pm$ \textbf{0.274} & N/A & \textbf{0.592} $\pm$ \textbf{0.295} & N/A \\\hline
  \end{tabular}
  \end{center}
\end{table}

\subsubsection{Static versus dynamic analysis}
\label{sec:static}
\begin{figure}[H]
  \begin{subfigure}[t]{0.5\linewidth}
    \begin{center}
      \includegraphics[width=0.5\linewidth]{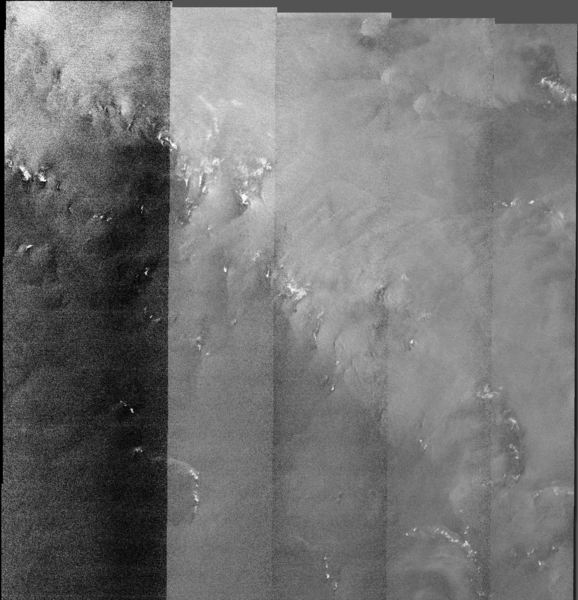}
      \end{center}
    \caption{Static scaling}
  \end{subfigure}
  \begin{subfigure}[t]{0.5\linewidth}
    \begin{center}
      \includegraphics[width=0.5\linewidth]{figures/dnoise2_downscale/figures/dnoise2/noint/indian/47.png}
    \end{center}
    \caption{Proposed dynamic scaling}
  \end{subfigure}
  \caption{An example of the importance of dynamically estimating
    scaling parameters for each image. In this case, the static scaling in
    this case overcompensates the true noise present in the
    image within the first swath.}
\label{fig:static_compare}
\end{figure}

While the overall effectiveness of the proposed method for denoising
the SAR images has been demonstrated, there still remains the
hypothesis that the ideal scaling parameters are different for each image.
This hypothesis was tested by computing the aggregate mean values of
$\hat{\mathbf{k}}$ with the proposed method for
HV ($\mathbf{\bar{k}} = [1.438,  0.942, 0.980,  1.010, 0.999]$) and VH
($\mathbf{\bar{k}} = [1.37, 0.932, 0.969, 0.993, 1.000]$) and
comparing the effect of scaling the noise field with
$\mathbf{\bar{k}}$ versus scaling the noise field by dynamically estimating $\mathbf{\hat{k}}$ for each image. 
\figref{fig:static_compare} shows an example of how using a static
estimate ($\mathbf{\bar{k}}$)
can provide sub-optimal denoising compared to using a dynamic estimate ($\mathbf{\hat{k}}$).

To test the significance of this effect, the same quantitative
experiment in section \ref{sec:exp_line}, which compared the NRMSE of the
linear regression with respect to the denoised measurement over ocean regions
between two different methods, was applied. Namely, the mean estimates for
$\mathbf{\bar{k}}$ were applied to scale the noise field as the
\textit{static} method versus scaling with $\mathbf{\hat{k}}$ using the \textbf{proposed}
method. Overall, the \textbf{proposed} method had
significantly lower NRMSE than the \textit{static} ($p = 5.8\times
10^{-3} < 0.05$) from a one-tailed t-test, thus supporting
our hypothesis that the ideal scaling factors are scene
independent. Once again, these results are summarized in Table \ref{tab:NRMSE_results}.

\subsubsection{IPF 3+}
\label{sec:ipf3}
\begin{figure}[H] 
  \begin{subfigure}{0.5\linewidth}
    \includegraphics[width=\linewidth]{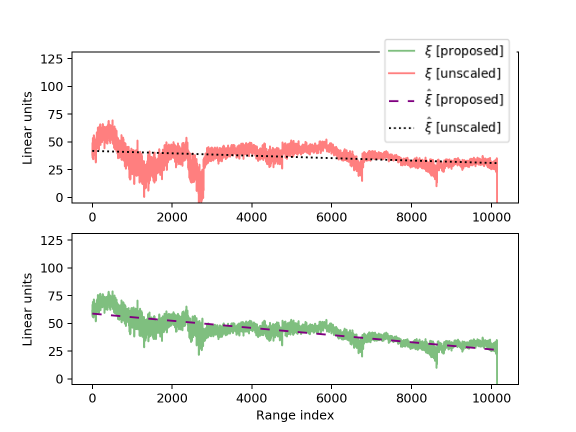} 
    \caption{Range plot and linear regression for ESA method (top) and the proposed method(bottom)}
  \end{subfigure}
  \begin{subfigure}{0.5\linewidth}
    \begin{minipage}[t]{\linewidth}
      \begin{tabular}{M{44mm}M{44mm}}
        ESA & Proposed \\        
        \includegraphics[width=\linewidth]{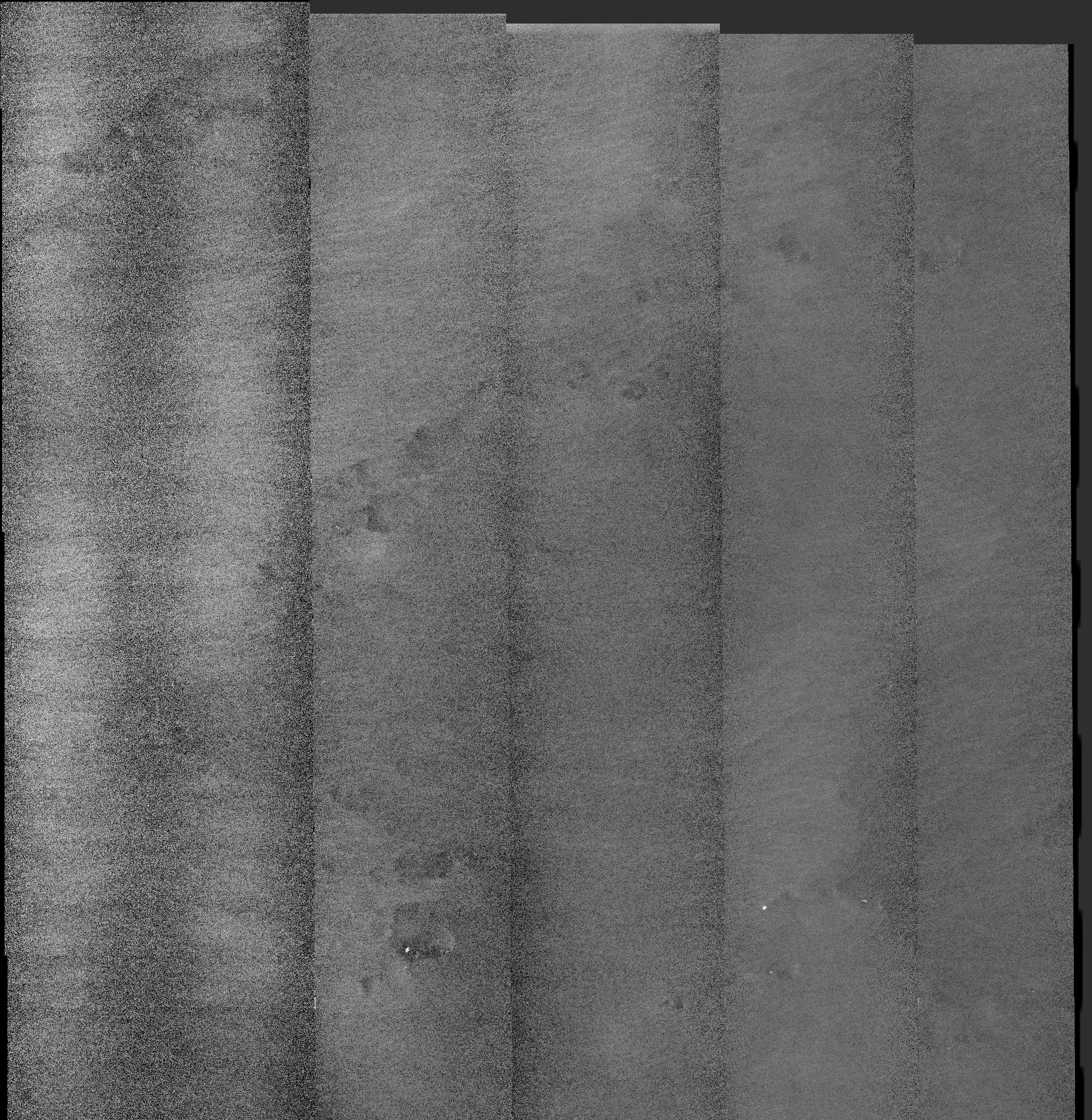} & \includegraphics[width=\linewidth]{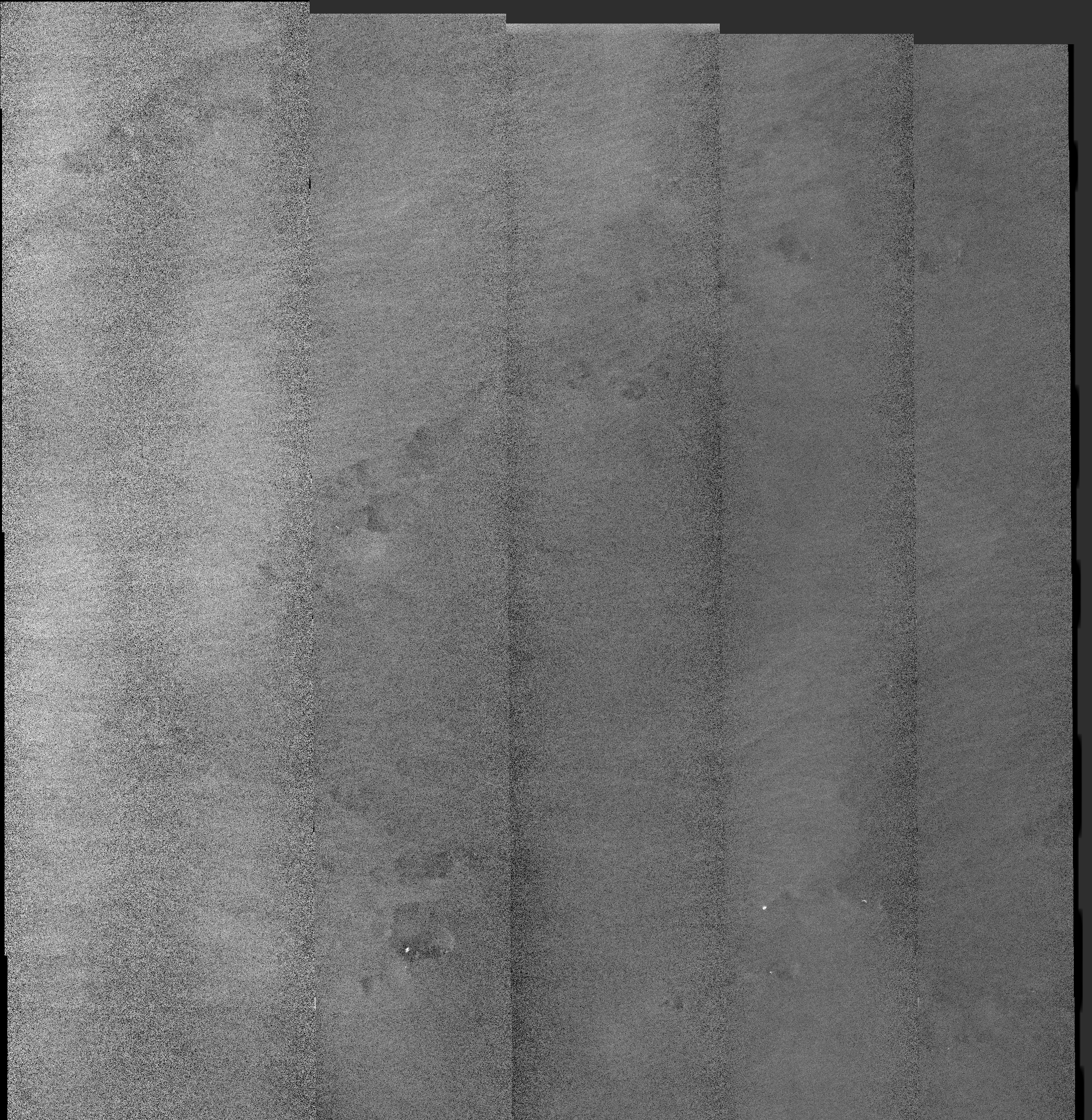} \\
      \end{tabular}
    \end{minipage}
    \caption{HV image results.}
  \end{subfigure}
  \caption{Brief comparison with IPF version 3+}
  \label{fig:ipf3_compare}
\end{figure}

With the changes from IPF version 3+, the generation of noise
  fields changed significantly. A new normalization scheme was added
  to help account for situations where the signal to noise ratio is
  low \citep{ipf3}. To evaluate the effect of the proposed method on
  these new noise fields, we collected 13, 12, 8, 14, and 9 images
  from the Arctic, Antarctic, Atlantic, Pacific, and Indian oceans
  with sensing dates after July 2019. We repeated the quantitative
  experiment with this data, while skipping the \textit{static}
  comparison for brevity. Table \ref{tab:NRMSE_results} indicate that
  the proposed method (NRMSE = 0.592) had significantly lower NRMSE
  than the ESA method (NRMSE = 0.685), with $p= 0.017 < 0.05$. As
  shown in Figure \ref{fig:ipf3_compare}, the improvement between the
  proposed method and the ESA method is milder than with IPF 2.91.

\section{Discussion}
\label{sec:discussion}
As mentioned in the introduction, the merit of the proposed
method is based on achieving four objectives. The first three objectives are
directly tied to the results of the two
experiments, while the last objective is achieved from the formulation
of the method itself.

The simulation experiment is based on the ideal scenario where our
assumptions of actual additive noise are true; in other words, the noise field is
linearly mis-scaled between subswaths and the scaling is variable among different
scenes. Under these circumstances, the proposed method provides an
almost perfect denoising of the selection of Sentinel-2 scenes
over coastal regions. In this scenario, the proposed method
produced significantly better denoising than the scaling from using the
static expected scaling parameters. This indicates that the proposed
method can effectively adapt to the dynamic-scaling of the noise field
in an ideal environment.

While the simulation experiment was a good demonstration that the
proposed method works well in a controlled setting where the ground
truth is known, obviously the characteristics of SAR and optical images
are inherently different and the experiment on SAR images is
required even if a ground truth reference is not known.
Overall, the proposed method made dramatic improvements on the visual quality of the SAR images. The method was able to estimate parameters in a wide
variety of backgrounds including those with heterogeneous textures,
such as sea-ice or cyclones. Compared to
subtracting an unscaled noise field, the proposed method suppresses the characteristics of the noise field to a much greater degree. Analytically, the evidence that the proposed method produces
significantly lower NRMSE over the range of ocean regions concurs to
the effectiveness of the proposed noise removal and the accomplishment
of the first two objectives.

The experiment on the SAR images also revealed the impact of
dynamically estimating scaling parameters. By using the mean
parameters generated by the proposed method as the static scaling
parameters, there are instances where the static parameters are
unsuitable for the subswaths. This
effect was significant in terms of NRMSE over range of ocean regions,
indicating that there is a clear benefit for dynamic estimation.

Also of note is that the proposed method requires little computational
overhead relative to the size of the noise field and image. The
computational complexity of constructing the linear system is of the
same order as a reduction along the azimuth, while solving the
$5\times5$ linear system of least-squares is trivial. The small size of the system is also beneficial
because a closed form solution can be used and no expensive iterative algorithms like the conjugate gradient method,
gradient descent, or grid search are required to estimate scaling
parameters. The low overhead is another advantage of the dynamic
parameter estimation aspect of the proposed method because the
parameters can be quickly estimated for each image and, unlike a
static estimation approach, requires no preparation in the sense of
collecting a training set.

The introduction of IPF 3+ greatly improved the quality of the
  noise field. In section \ref{sec:ipf3}, it was demonstrated that the
  proposed method still provided some modest improvement, but the
  visual differences can be subtle. However, archived Sentinel-1 data
  is not updated with the newest IPF and are still susceptible to the
  mis-scale from IPF 2.91 and below. Therefore, our method provides
  a way for improving quality that can be applied to both older
  and current Sentinel-1 data.

\begin{figure}[H]
  \begin{subfigure}[t]{0.48\linewidth}
  \includegraphics[width=\linewidth]{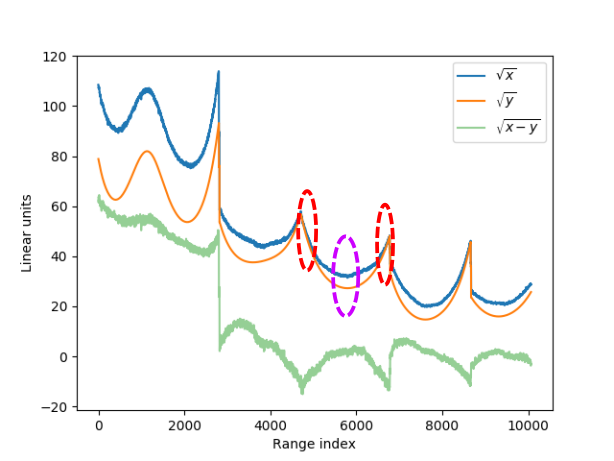}
  \caption{ID: \small{20180930T081301\textunderscore
    20180930T081401\textunderscore 023925\textunderscore 029CB0} The
    relative difference between the measurements and the noise is
    greater in the centre of subswaths (purple) 
    than the 
  right and left extremities (red) within the same subswath.}
  \label{fig:umisfit_exmp}
\end{subfigure}\hfill
\begin{subfigure}[t]{0.48\linewidth}
  \includegraphics[width=\linewidth]{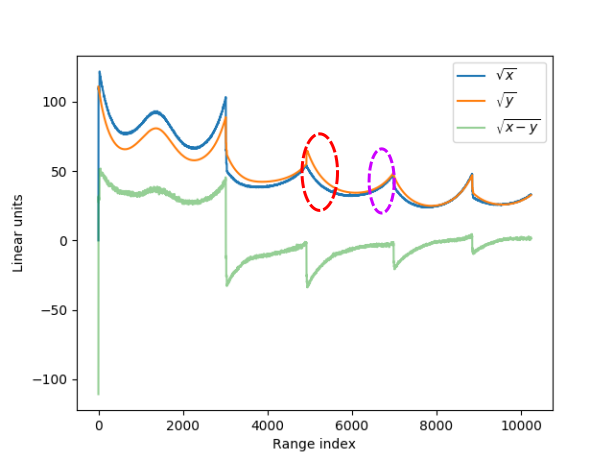}
  \caption{ID: \small{20190322T212224\textunderscore 20190322T212324\textunderscore 015472\textunderscore 01CFB5:} The relative
  difference between the noise and signal is different for the right (red)
  and left extremities (purple) within the same subswath. 
  }
\label{fig:leftmisfit_exmp}
\end{subfigure}
\begin{subfigure}[t]{0.48\linewidth}
  \includegraphics[width=\linewidth]{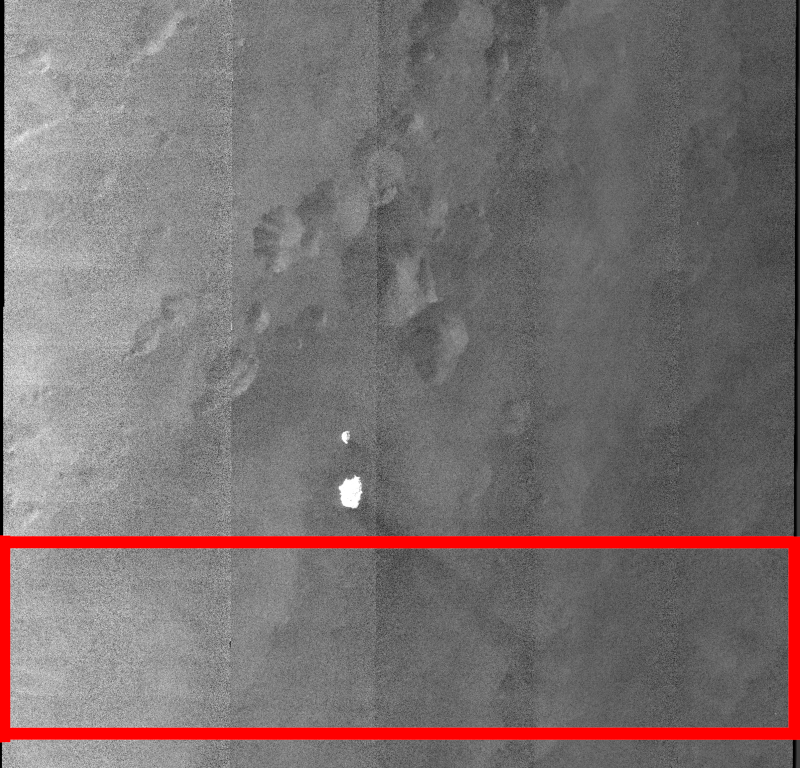}
  \caption{ID: \small{20180930T081301\textunderscore 20180930T081401\textunderscore 023925\textunderscore 029CB0:} Input image with evaluation range highlighted in red.}
\end{subfigure}\hfill
\begin{subfigure}[t]{0.48\linewidth}
  \includegraphics[width=\linewidth]{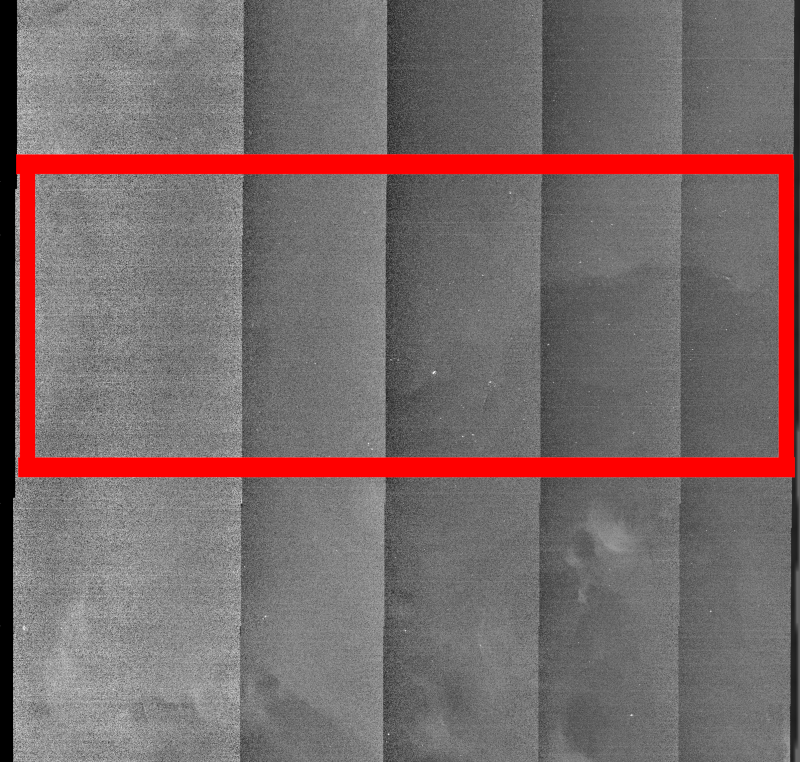}
  \caption{ID: \small{20190322T212224\textunderscore 20190322T212324\textunderscore 015472\textunderscore 01CFB5:} Input image with evaluation range highlighted in red.}
\end{subfigure}
\caption{ Mean signal
    and noise with respect to range over an ocean sub-region. In both
    examples the noise
    is not fit properly for subswaths EW2, EW3, EW4, and
    EW5.}
\end{figure}

Despite the aforementioned success, some issues still exist in the SAR-images. 
As mentioned in \secref{sec:exp_line}, some images have discontinuities
between adjacent subswaths even after denoising. These
issues do not seem to be from an incorrect estimation of scaling
parameters. Rather, they seem to be caused from the shape
of the ESA noise field not always fitting to the actual shape of noise imposed on
the image, particularly in the range
direction. An extreme example of this is shown in \figref{fig:umisfit_exmp}, where the left and right extremities of subswaths EW2, EW3, EW4, EW5 have a higher relative compensation than the centre of the
swath. Another example in \figref{fig:leftmisfit_exmp} shows a different style of misfit, where the left extremities of the subswaths have higher
relative compensation compared to the right extremity. This
shows that for some images, the provided noise fields are
calibrated incorrectly in a more complex manner than linear scale. Consequently, this misfit accounts for errors in the noise removal that
cannot be compensated by scaling individual subswaths alone. Correcting this will require
correcting the shape of the noise curves either through enhanced
calibration or by creating a more flexible empirical model to account for the
shape of the curves. This is consequently an area of study
for future work. While it would be interesting to adapt these methods to co-polarized images or images from Sentinel-1 modes such as IW, we suspect that the higher signal-to-noise ratios from these methods would make application of our methods to these types of images less necessary.

\section{Conclusion}
We created a quadratic objective function to model the
characteristics of the estimated noise field in TOPSAR cross-polarized
images. Our method uses this objective function to estimate scaling parameters for each of the
subswaths in a scene. The method compared favourably to unscaled noise removal in both a
simulation experiment on non-SAR images on an experiment selection of Sentinel-1
cross polarized TOPSAR images over major ocean-divisions around the
world. These experiments showed that the algorithm works on a
variety of backgrounds, has no requirements of acquiring a training
data set, provides dynamic scaling parameter estimation for each image,
and has low computational requirements. These merits are beneficial for any practitioner
who uses cross-polarized Sentinel-1 EW images, as the algorithm can be
conveniently applied with little preparation required.

\section*{Conflict of Interest}
Declarations of Interest: None

\section*{Acknowledgments}
We acknowledge the funding support of the Natural Sciences and Engineering Research Council of Canada (NSERC) (RGPIN-2017-04869, DGDND-2017-00078, RGPAS-2017-50794,
RGPIN-2019-06744) and from the University of Waterloo.

Thanks are also extended to Prof. K. Andrea Scott, Mohsen Ghanbari, and Mingzhe Jiang for their feedback on the manuscript.\\

\appendix
\section{Inner Product Formulations}

\begin{claim} $L^{\sim} \in \{L^{A}, L^{B}, L^{R}\}$ can be
  represented with an arbitrary inner product formulation
  $[\mathbf{v}^{\sim} -
  \mathbf{C}^{\sim}\mathbf{\hat{k}}]^T[\mathbf{v}^{\sim} -
  \mathbf{C}^{\sim}\mathbf{\hat{k}}]$, using some vector
  $\mathbf{v}^{\sim}$ and matrix $\mathbf{C}^{\sim}$.
  \end{claim}

\begin{proof}
First, discard the connotations associated with the variables $i$,
$j$, and $n$ in the main text. Without loss of generality, let
$L^{\sim} = \sum\limits^N_{i=1}  [w(i)[\hat{\phi}_{a(2i-1)}^{2i-1} -
\hat{\phi}_{a(2i)}^{2i}]]^2$ given the linear denoising model $\hat{\phi}_{a(j)}^{j} = x_{a(j)}^{j} - \hat{k}_{a(j)}y_{a(j)}^{j}$, with $x_{a(j)}^j$ representing a selection of the measurement, $y_{a(j)}^j$ representing a selection of the estimated noise, and a subswath $a(j) \in \mathcal{A}$. Then for any $i$,
\begin{equation}
  \begin{aligned}
    {w(i)} [\hat{\phi}_{a(2i-1)}^{2i-1} - \hat{\phi}_{a(2i)}^{2i}] &= {w(i)}[x_{a(2i-1)}^{2i-1} - \hat{k}_{a(2i-1)}y_{a(2i-1)}^{2i-1}] - [y_{a(2i)}^{2i} - \hat{k}_{a(2i)}y_{a(2i)}^{2i}] \\
    &= {w(i)}[x_{a(2i-1)}^{2i-1} - x_{a(2i)}^{2i} - [\hat{k}_{a(2i-1)}y_{a(2i-1)}^{2i-1} - \hat{k}_{a(2i)}y_{a(2i)}^{2i}]].
  \end{aligned}
  \label{eq:layout}
\end{equation}
We can encode (\ref{eq:layout}) for all $i$ into a vector $\mathbf{v}^\sim - \mathbf{C}^{\sim}\mathbf{\hat{k}}$ where $\mathbf{v}^\sim$ contains the terms with $x$ and $\mathbf{C}^\sim$ contains the terms with $y$. More precisely,
\begin{equation}
  \mathbf{v}^\sim \in \mathbb{R}^N = \begin{vmatrix}
    {w(1)} [x_{a(1)}^1  - x_{a(2)}^2] \\
    \vdots \\
    {w(i)} [x_{a(2i-1)}^{2i-1} - x_{a(2i)}^{2i}]\\
      \vdots \\
      {w(N)} [x_{a(2N-1)}^{2N-1} - x_{a(2N)}^{2N}] \\
      \end{vmatrix}
    \end{equation}
    and
    \begin{equation}
      \mathbf{C}^\sim (i,n) = \left\{\begin{array}{lr}
                                                          {w(i)} [y_{a(2i-1)}^{2i-1} - y_{a(2i)}^{2i}] & \text{if } a(2i-1) = a(2i) = n \\
                                                         {w(i)}y_{a(2i-1)}^{2i-1} & \text{if } a(2i-1) = n \text{ and } a(2i) \not{=} n \\
                                                         -{w(i)}y_{a(2i)}^{2i} & \text{if } a(2i) = n \text{ and } a(2i-1) \not{=} n \\
                                                         0 & \text{otherwise}
                                                             \end{array} \right\}, 
    \end{equation}
    with the index $n$ representing the different subswaths $\in \mathcal{A}$ and $\mathbf{C}^\sim \in \mathbb{R}^{N\times5}$.
    Then by the rules of matrix multiplication
    \begin{equation}
      \sum^N_{i=1} [w(i) [\hat{\phi}_{a(2i-1)}^{2i-1} - \hat{\phi}_{a(2i)}^{2i}]]^2 =  [\mathbf{v}^{\sim} - \mathbf{C}^{\sim}\mathbf{\hat{k}}]^T[\mathbf{v}^{\sim} - \mathbf{C}^{\sim}\mathbf{\hat{k}}]
    \end{equation}
\end{proof}

\bibliographystyle{model-names2.bst}\biboptions{authoryear}
\raggedright
\bibliography{main}

\end{document}